\documentclass[10pt,journal,letterpaper,compsoc]{IEEEtran}
\pdfoutput=1

\ifCLASSOPTIONcompsoc
\else
\fi

  \usepackage[dvips]{graphicx}
  \DeclareGraphicsExtensions{.eps}
  \usepackage[update,prepend]{epstopdf}

\usepackage{amssymb}
\usepackage[cmex10,fleqn]{amsmath}
\usepackage{longtable}
\usepackage{booktabs}	

\usepackage{graphicx}
\usepackage{times}

\usepackage{subfigure}

\usepackage{verbatim}

\usepackage[pdftex,bookmarks]{hyperref}
\hypersetup{
  pdfauthor = {Steven Bergner},
  pdftitle = {Paraglide: Interactive Parameter Space Partitioning for Computer Simulations},
  pdfsubject = {Cluster by similarity in application dependent objective measures to discretize continuous parameter spaces.},
  pdfkeywords = {multi-dimensional sampling, integration, approximation, visualization},
  colorlinks=true,
  linkcolor= black,
  citecolor= black,
  pageanchor=true,
  urlcolor = black,
  plainpages = false,
  linktocpage
}

\usepackage[usenames,dvipsnames]{color}
\newcommand{\sbnote}[1]{{\color{BurntOrange}[*** SB: #1 ]}}

\newcommand{\bluenote}[1]{{\color{RoyalBlue}[[*** #1 ]]}}

\newcommand{\refn}[1]{#1\index{#1}} % create a reference in the index
\newcommand{\df}[1]{{\em #1}}  % define a term..
\newcommand{\defn}[1]{\df{\refn{#1}}} % ..and include a reference

\newcommand{\cname}[1]{{\small \tt #1}} % to name components in the code
\newcommand{\sname}[1]{{\small \sf #1}} % formatting for the name of a software or system
\newcommand{\R}[1]{{\small \textsf{\textbf{R#1}}}} % referring to a requirement

\newcommand{\SC}[1]{Section~\ref{#1}}

\newcommand{\FG}[1]{Figure~\ref{#1}}

\newcommand \parao[1] {\vspace{.25em} {\setlength{\parindent}{0pt} \bf #1}}
\newcommand \para[1] {\parao{#1:}}
\newenvironment{tight_itemize}{
\begin{itemize}
  \setlength{\itemsep}{1pt}
  \setlength{\parskip}{0pt}
  \setlength{\parsep}{0pt}}{\end{itemize}
}

\newcommand \mb[1]{\mathbf #1}

\newcommand \RR{\mathbb{R}}

\usepackage{bbold}
\newcommand \one{\mathbb{1}}

\providecommand{\abs}[1]{\left|#1\right|}
\providecommand{\norm}[1]{\lVert#1\rVert}
\newcommand{\vol}{\mathsf{vol}}

\newcommand{\Var}[1]{\mathsf{V}\left[#1\right]}

\usepackage{ifthen}
\newboolean{optional_content}
\setboolean{optional_content}{false} % true or false indicates whether or not to show optional content.
\begin{document}

% A Visual Interface for User Guided Exploration and Optimization of Computer Simulations
\newcommand \titletext {Paraglide: Interactive Parameter Space Partitioning for Computer Simulations}

\title{\titletext}

\author{
%Steven~Bergner, Michael~Sedlmair, Sareh~Nabi-Abdolyousefi, Ahmed~Saad, Torsten~M{\"o}ller
\begin{tabular}{ccccc}
	Steven~Bergner & Michael~Sedlmair & Sareh~Nabi-Abdolyousefi & Ahmed~Saad & Torsten~M{\"o}ller \\
	\scriptsize sbergner@cs.sfu.ca &
	\scriptsize msedl@cs.ubc.ca &
	\scriptsize sna39@sfu.ca &
	\scriptsize aasaad@cs.sfu.ca &
    \scriptsize torsten@sfu.ca
\end{tabular}
%
%%Anno~Nynmous et al.,~\IEEEmembership{Student Member,~IEEE}% ,
%%        John~Doe,~\IEEEmembership{Fellow,~OSA,}
%%        and~Jane~Doe,~\IEEEmembership{Life~Fellow,~IEEE}%
\IEEEcompsocitemizethanks{\IEEEcompsocthanksitem S.~Bergner, A.~Saad, and T.~M{\"o}ller are with the Dept.
of Computing Science, Simon Fraser University (SFU), Burnaby, BC, V5A 1S6. %\protect\\
% note need leading \protect in front of \\ to get a newline within \thanks as
% \\ is fragile and will error, could use \hfil\break instead.
\IEEEcompsocthanksitem M.~Sedlmair is with Imager Lab, University of British Columbia. %
\IEEEcompsocthanksitem S.~Nabi-Abdolyousefi is with the Dept. of Applied Mathematics, SFU.}%
\thanks{}%
}

% The paper headers
%\markboth{IEEE TVCG submission draft}%
\markboth{Technical Report SFU-CMPT TR 2011-06}%
{Bergner \MakeLowercase{\textit{et al.}}: \titletext}

\IEEEcompsoctitleabstractindextext{%
\begin{abstract}
In this paper we introduce \sname{paraglide}, a visualization system designed for interactive exploration of parameter spaces of multi-variate simulation models. 
To get the right parameter configuration, model developers frequently have to go back and forth between setting parameters and qualitatively judging the outcomes of their model.
During this process, they build up a grounded understanding of the parameter effects in order to pick the right setting.
Current state-of-the-art tools and practices, however, fail to provide a systematic way of exploring these parameter spaces, making informed decisions about parameter settings a tedious and workload-intensive task.
\sname{Paraglide} %, the system we propose, 
endeavors to overcome this shortcoming by 
assisting the sampling of the parameter space and the discovery of qualitatively different model outcomes.
This results in a decomposition of the model parameter space into regions of distinct behaviour.
%To address practical questions in different use cases, we identify the frequent task of constructing meaningful features (scalar and vector) that may be directly used to enhance a visualization or that lead to derived similarity or distance measures for further processing.
%Working with interpreter based back-ends, such as Matlab, R, or Python, our system supports the interactive construction of feature variables.
We developed \sname{paraglide} in close collaboration with experts from three different domains, who all were involved in developing new models for their domain. We first analyzed current practices of six domain experts and derived a set of design requirements, then engaged in a longitudinal user-centered design process, and finally conducted three in-depth case studies underlining the usefulness of our approach.
\end{abstract}

% may also check http://www.acm.org/class/1998/

%% Note that keywords are not normally used for peerreview papers.
%\begin{IEEEkeywords}
%Computer Society, IEEEtran, journal, \LaTeX, paper, template.
%\end{IEEEkeywords}
}

% \teaser{
%  \includegraphics[width=\linewidth]{eg_new}
%  \centering
%   \caption{New EG Logo}
% \label{fig:teaser}

% make the title area
\maketitle

%\vspace{-2em}

% To allow for easy dual compilation without having to reenter the
% abstract/keywords data, the \IEEEcompsoctitleabstractindextext text will
% not be used in maketitle, but will appear (i.e., to be "transported")
% here as \IEEEdisplaynotcompsoctitleabstractindextext when compsoc mode
% is not selected <OR> if conference mode is selected - because compsoc
% conference papers position the abstract like regular (non-compsoc)
% papers do!
\IEEEdisplaynotcompsoctitleabstractindextext
% \IEEEdisplaynotcompsoctitleabstractindextext has no effect when using
% compsoc under a non-conference mode.

% For peer review papers, you can put extra information on the cover
% page as needed:
% \ifCLASSOPTIONpeerreview
% \begin{center} \bfseries EDICS Category: 3-BBND \end{center}
% \fi
%
% For peerreview papers, this IEEEtran command inserts a page break and
% creates the second title. It will be ignored for other modes.
\IEEEpeerreviewmaketitle

\begin{comment}
argumentative trail head:
- motivate modelling: real world and formal systems, each having their own mechanism of entailment, e.g. physical law or logical inference. Human intervention is required to ensure correspondence among the different systems, at least at the outset during model construction and possibly later controlling adjustments as more observations are made when running the model.
- types of observations: 
 field data: measurements at different levels of organization 
 			 (velocity of a particle, temperature of a fluid), 
 computed data: synthetic data generated by a computational model,
 derived properties: quantities obtained by studying formal properties of the model,
 user input: manually assigned feature variables, labels etc.
- parameter space partitioning: 
  provide qualitative data that allows to compare different model versions.
- a model is defined as a set of relations among variables for which it is possible to compute 
\end{comment}

%-------------------------------------------------------------------------
%\section{Visual Inspection of Computer \\Simulation Models}
\section{Linking Formal and Real Systems}

\IEEEPARstart{A}{t} the heart of computational science is the simulation of real-world scenarios. 
%Computational power today enables researchers to build more complex models that may involve large numbers of variables and relations among them. 
%
%During modelling, different forces acting on a closed system are calibrated using parameters, that - when set correctly - will allow the model to present a specific behaviour. 
%
%This could be a common, natural behavior, an extreme or dangerous behavior as well as an artifical behavior, not occurring in reality.% 
As it becomes possible to mimic increasingly comprehensive effects, it remains crucial to ensure a close correspondence between formal model and real system in order to draw any practically relevant conclusions.
%
%However, the process of understanding a system goes through different phases with different levels of detail at which relationships among variables can be defined. At early stages, reliable measurements may not be available for all phenomena that people have witnessed about the system. 
%% This especially may be the case for effects that occur at higher levels of organization in complex systems.
%In this case, it is still possible to construct a hypothetical model, visualize the data it generates and inspect, whether the qualitative aspects of the results fit with known experience.
%
%% The following are some more details illustrating the optimization workflow.
%It might be possible to compare multiple different settings by some chosen domain specific performance measure. In cases, where this measure can be expressed symbolically, one might optimize it algebraically. If it can be computed, an iterative numerical optimization is possible. However, if the distinction between good or bad outcome of the computation depends on the assessment by a human domain expert or is subjective for other reasons, then a systematic approach to the visual exploration process is highly desirable.
%
A well-established practical problem in this setting is the calibration of good parameter configurations that strengthen the fitness of the model~\cite[Ch.~1]{Saltelli:2008:gsa}.
Even after matching model output with measured field data, there may still be free parameters that can be controlled to adjust the behaviour of the computer simulation.
This can happen, if the expressive power
of the model exceeds the number of available measurements, or if the measurements are so noisy that several different model instances are equally acceptable.
%
\begin{comment}
To formally address this case, it is possible to introduce additional regularizing criteria that a solution has to fulfil.
Beyond that, 
the user could be given a method to interactively tune free parameters of the model to favour more plausible solutions that match prior experience. 
\end{comment}
%
In such a case, a domain expert could be involved to interactively tune free parameters of the model in order to favour solutions that match prior experience, theoretical insight, or intuition.

%Despite the continuous nature of its input parameter space, it is possible that for certain parameter changes a model drastically changes its behaviour as represented by its set of dependent output variables.
%To ensure that not too much time is wasted, when searching for parameter choices that lead to plausible model behaviour, a systematic approach during the adjustment is highly desirable.

Towards that goal, we recognize that the {\em optimization} of parameters for some notion of performance is distinct from the objective to {\em discover regions} in parameter space that exhibit qualitatively different system behaviour, such as fluid vs. gaseous state, or formation of various movement patterns in a swarm simulation.
Optimization is one focus of statistical methods in experimental design and has great potential for integration with visual tools, as for instance demonstrated recently by Torsney-Weir et al.~\cite{Torsney-Weir:2011:tuner}.
The focus of this paper is on the latter aspect of qualitative discovery. 
This can support the understanding of the studied system, strengthen confidence in the suitability of the modelling mechanisms and, thus, become a substantial aid in the research process. 

% local sensitivity analysis: How do parameter changes affect changes in output?
% global SA: What are the regions in the parameter domain that lead to output that does not change much.

In the context of modelling this is a novel viewpoint, since typical approaches calibrate one best version of the model and then study how it behaves. 
%Hence, we start by confirming that there is a practical need.
To put regional parameter space exploration into practice, a number of challenges have to be overcome.
To identify and address those, we (a) performed a field analysis of three application domains and derived a list of requirements, (b) present \sname{Paraglide}, a system that addresses these requirements with a set of interaction and visualization techniques novel for this kind of application area, (c) conducted a longitudinal field evaluation of \sname{Paraglide} showing practical benefits. In summary, \sname{Paraglide} sets out to make the following contributions to computational modelling:
\begin{itemize}
\item Parameter region construction is promoted as a separate user interaction step during experimental design. This allows to address different efficiency issues of multi-dimensional sampling.
\item A common step in explorative hypothesis formation is the construction of additional dependent feature variables and goal functions. \sname{Paraglide} facilitates this with interpreter based back-ends. 
% that allow for runtime adjustments to the code.
Also, this seamlessly integrates model code from sources such as MATLAB, R, or Python. 
\item Qualitatively distinct solutions are identified and the parameter space of the model is partitioned into the corresponding regions. This allows to visually derive global statements about the sensitivity of the model to parameter changes, which traditionally is studied locally.
\end{itemize}

%\item Complex models of a large number of variables may require a substantial number of sample points to capture a sufficient amount of detail. Hence, the experimental design process that decides which parameter configurations to compute model output for, is decomposed into several phases --- allowing to individually address different efficiency issues. 
%\item A common step in explorative hypothesis formation is the construction of additional dependent feature variables or goal functions. 
%We facilitate this with interpreter based back-ends, allowing for runtime adjustments to the code.
%\item We facilitate an interactive partitioning of the parameter space into regions of distinct system behaviour. This also allows to visually derive statements about the sensitivity of the model to parameter changes on a more global scale.
%\item The practicality of the system is shown by qualitative evaluation with different target user groups. 

%Are a few words about online (steering) vs offline (sampling) parameter manipulation, and a note on optimization vs. discovery in order or can we postpone that to \SC{sec:relwork}?}

% -----------------------------------------------------------------------
\section{Domain Characterization \label{sec:requirements} }
%and \\ Requirement Analysis }

In order to get a more detailed understanding of needs and requirements, we engaged in 
a problem characterization phase. We conducted contextual interviews with six experts from three different domains: engineering, mathematical modelling, and segmentation algorithm development. 
In the following, we characterize the investigated domains. 
Based on that, we summarize design requirements in \SC{sec:problem} that are more general yet grounded in real-world application areas. %\FG{tab:requirements}. 

\subsection{Mathematical Modelling: Collective behaviour in biological aggregations\label{sec:swarms}}
\begin{comment}
Techniques coming to bear in this case:
- overview of distribution of initial set of configurations (Raluca's cases)
- iterative exploration of regions of the paramter space
\end{comment}

% What
Our first target group are two researchers studying properties of a mathematical model that describes biological
aggregations.
% Eulerian models use partial differential equations and track the density of the individuals.
% Mathematical models help us gain a better understanding and a clearer explanation for the
% observed group behaviours in nature. 
% 
% Flocks of birds, swarms of insects, schooling behavior of fish, herds of quadrupeds, bacterial swarms, and many other biological   
% aggregations have attracted the attention of scientists for a long time.
% 
% %Lagrangian models are individual-based models that study the collective behavior of the
% animal groups. Most of these models consider the three different interactions among group
% members, that is, attraction towards distant individuals, repulsion towards the close ones
% and a tendency to align with neighbors
% 
%
% Why / Goal
Furthering the understanding of such spatial and  spatio-temporal patterns helps, for instance, to better predict animal migration behaviour. 
Modelled patterns can inform measures to contain plagues of locusts and positively affect quality of life in third world countries~\cite{Buhl:2006:disorder}. It may also help to better understand how, where, and when fish aggregations form and contribute to more efficient fishing strategies
\ifthenelse {\boolean{optional_content}}{
\cite{Parrish:ube,Radakov}.
}{\cite{Parrish:ube}.} %end of optional_content 
 %shows how increasing knowledge of how, where, and when 
%fish aggregations form,
%or understanding the cause of schooling behaviour can help fishermen to catch fish more
%efficiently 

%Recently, engineers have shown interest in swarms, too. Understanding
%the way individuals communicate with each other and make coordinated decisions can help
%engineers to develop automated systems such as remote-controlled vehicles and autonomous
%multi-robot teams 
%\ifthenelse {\boolean{optional_content}}{
%\cite{Balch:Arkin, Gazi:Passino, Giulietti:etal,Kube:Hong}. 
%}{\cite{Gazi:Passino, Giulietti:etal}.} %end of optional_content 

% What exactly?
To study 
%\bluenote{MS: What's the goal, studying or predicting?} 
%\sbnote{For now: study them to see which ones are present and how they correspond to real biological aggregations.}
those spatio-temporal patterns, our participants developed a mathematical model~\cite{Fetecau:2009:biogroups,Lutscher:Stevens}
consisting of a system of partial differential equations (PDEs) that express in one spatial dimension how left and right travelling densities of individuals move and turn over time.
More details are given in the chapter notes.
The basic idea is to take three kinds of social forces into account --- namely attraction, repulsion, and alignment --- that act globally among the densities of individuals. 
Attraction is the tendency between distant individuals to get closer to each other, repulsion is the social force that causes individuals in close proximity to repel from each other, and alignment represents the tendency to sync the direction of motion with neighbours.
Solving the model for different choices of coefficients produces many complex spatial and spatio-temporal patterns observed in nature,
such as stationary aggregations formed by resting animals, zigzagging flocks of birds, milling schools of fish, and rippling behaviour observed 
in Myxobacteria. %, in which right and left traveling ridges of high cell densities pass through each other.% \cite{Lutscher:Stevens}.

Our use case is part of a Master's thesis on this subject with a focus on comparing two versions of their model. In the first one the velocity is 
constant. In the second one the individuals speed up or slow down as a response
to their social interactions with neighbours. 
%
% Problem statement:
%The primary modelling interest is in comparing the spatial patterns obtained as solutions of the two PDE systems. This 
Comparing these models requires to solve them numerically for several different configurations. Each one of them corresponds to one specific choice of the $14$ model parameters, including the coefficients for the three postulated social forces.
The output of the simulation is a spatio-temporal pattern of population densities. The number of basis functions that gives the resolution in space and time can be chosen to adjust the accuracy/runtime trade-off between $2$ minutes and an hour.
With $5$ minutes each, one can perform a full computation of close to $300$ sample points in the duration of a single day.

% How / Current practices
To better understand the space of possible solutions, 
% predictive power of their model \bluenote{MS:is this their main goal?}, 
our participants manually explored the parameter combinations of their model and 
demonstrated its capability to reproduce a variety of complex patterns. %, which may also depend on the initial population density.
While it is difficult to classify all possible patterns, there are a few standard solutions among them, for which established analysis techniques exist. 
%
%
%% METHOD OF LINEAR STABILITY ANALYSIS
% Moved some of the description to 5.1
In particular, they focus on the solutions of the system that do not change over time and space --- so called spatially homogeneous steady states. 
A linear stability analysis of these steady states results in negative or positive growth rates for different perturbation frequencies, which respectively indicate stable and unstable solutions. 
\begin{comment}
Also, we are interested in 
finding the stability of these solutions numerically and analytically and comparing 
the results for the constant and non-constant velocity models. The comparison 
helps us to have a better understanding of the effect of the velocity on the behavior 
of the individuals. Also, helps to decide on a model that gives a better description 
of what is happening in nature.
\end{comment}
%
%Choosing configurations manually is a time consuming task and also makes comparison of varying conditions more difficult. 
There is a hypothesized relationship between the stability of steady states and the potential for pattern formation. This leads to a derived, more specific goal of the study.
In particular, it allows to compare constant and non-constant velocity models by inspecting the change in shape of the parameter regions that lead to (un-)stable steady states.
% Writing for-loops in Matlab scripts is current practice. The factorial designs this leads to are inefficient in a multi-dimensional setting.
%
%% Problems
%There are several free parameters that can be adjusted to configure the forces, where slightly different configurations may lead to drastically different behaviour \bluenote{MS:is this the big problem?} Due to their complexity, visual judgement is required to categorize the patterns. 
%
%%In different animal groups it is possible to observe common patterns of movement, such as stationary aggregations formed by resting animals, migrating herds, zigzagging flocks of birds, and milling schools of fish.
%%Understanding these patterns may enable us \bluenote{MS:us?} to better predict animal migration behaviour. As a specific example, this could have impact on containment of plagues of locusts and positively affect quality of life in third world countries.
%
A discussion on how 
\sname{paraglide} affects our participant's workflow is given in \SC{sec:swarmsres}.

\begin{comment}
main goal was to compare non-constant and constant velocity.
speed of the transition is interesting feature for this comparison, too.
% discussion similar to section 4.2.1
large values of $q_a$ introduce instability, because the density distribution tends to concentrate, rather than stay flat (e.g. spatially homogeneous).
$q_r$ is a parameter that increases stability. 
reading the graph to interpret how it changes
\end{comment}

%In different animal groups it is possible to observe common patterns of movement, such as stationary aggregations formed by resting animals, migrating herds, zigzagging flocks of birds, and milling schools of fish.
%Understanding these patterns may enable researches to better predict animal migration behaviour. As a specific example, this could have impact on containment of plagues of locusts and positively affect quality of life in third world countries.

\begin{figure}[hb]
\centering
\begin{tabular}{lr}
  \subfigure[raw data]{\includegraphics[height=3cm]{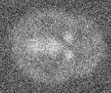}} &
  \subfigure[ground truth]{\includegraphics[height=3cm]{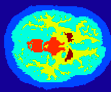}} \\

% Should we also show the source image that is input to the algorithm?
% These are from the n5_50 data set:
%'count'	'alpha1'	'alpha2'	'alpha3'	'alpha4'	'alpha5'	'alpha6'	'alpha7'	'segma'	'dice2'	'dice3'	'dice4'	'dice5'	'dice6'	  
%  'Error2'	'Error3'	'Error4'	'Error5'	'Error6'
%13	0.3668390 3.595200 0.4586530 0.8067480 0.5448060 1.866313 1.553262 0.6444850 0.9776800 0.8959470 0.7645130 0.9068010 0.8135590 
%  0.0001590000 0.01940000 0.007863000 0.05229200 0.066896
  \subfigure[config\# 13, dice$_6=.8136$]{\includegraphics[height=3cm]{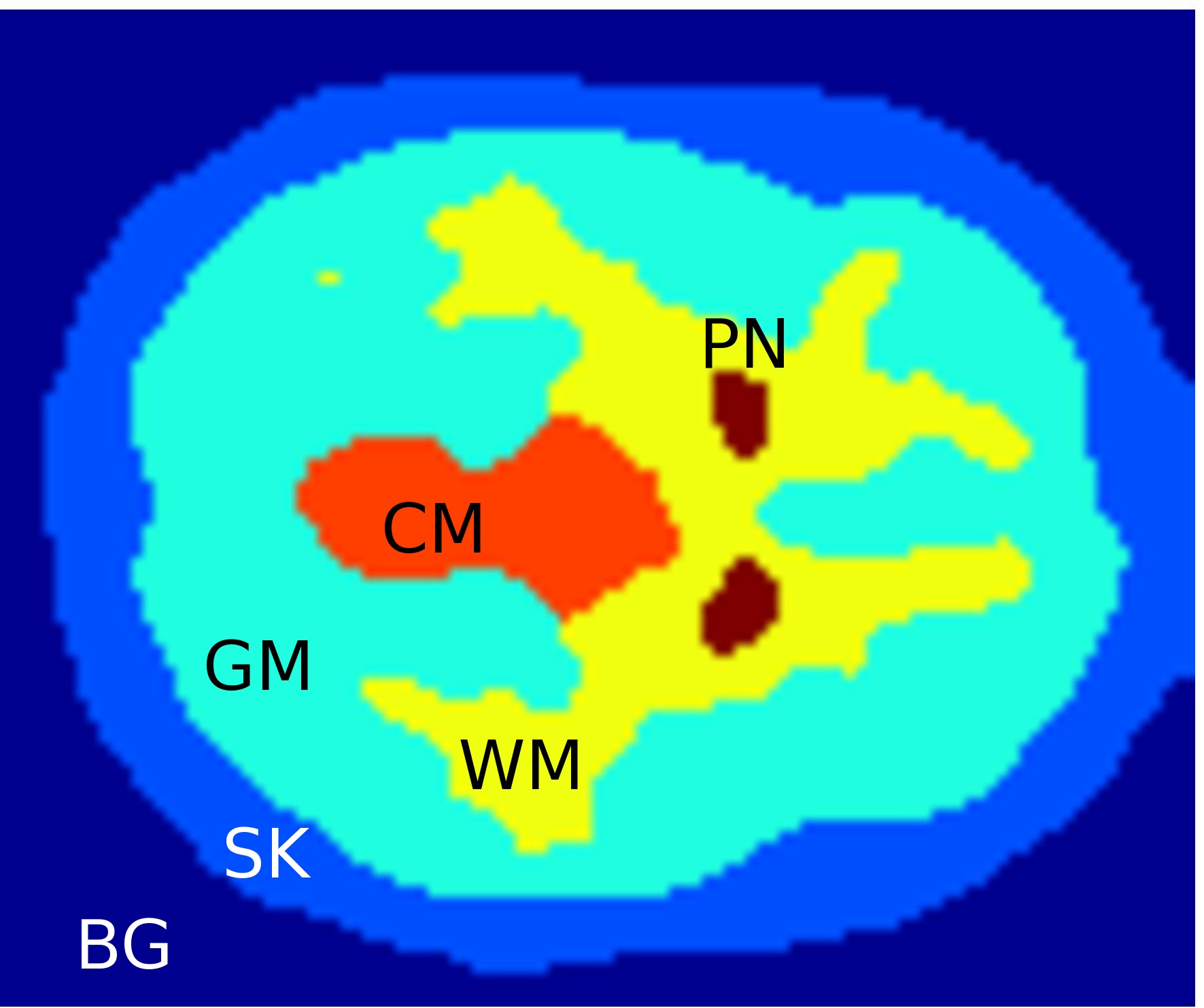}} % count: 13, dice6: .81
%44	0.8775920 0.2128050 0.1222200 0.3448360 1.630549 1.112137 0.1662540 0.7294980 0.9784210 0.8917480 0.7752440 0.8511110 0.8127850 
%  0.003302000 0.01558300 0.004692000 0.04892600 0.06351
  & \subfigure[config\# 44, dice$_6=.8128$]{\includegraphics[height=3cm]{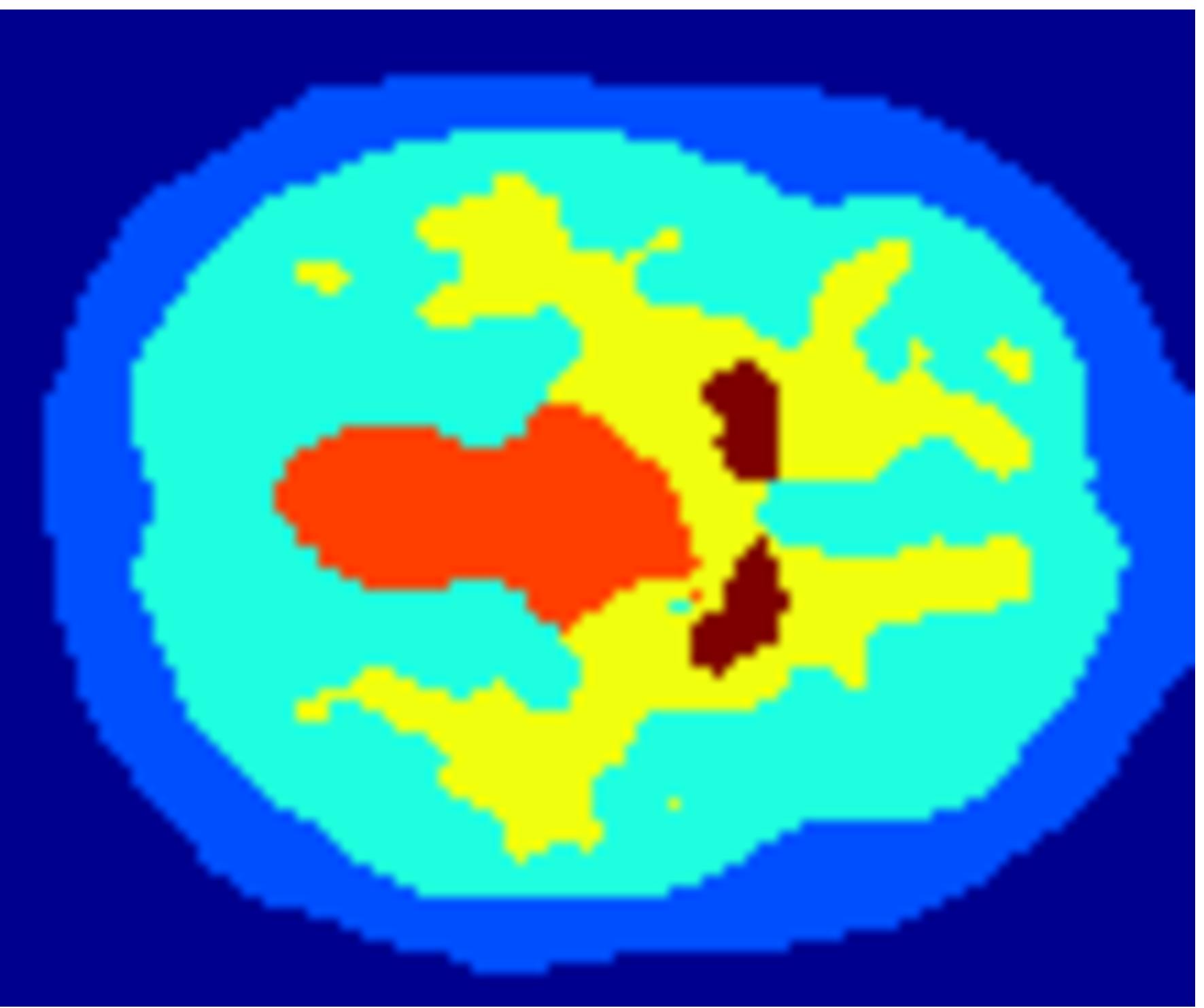}} % count: 44, dice6: .81
%%37 0.8159140 0.5116010 0.2082620 0.5606460 0.6754980 1.463223 1.477655 0.1025930 0.9456520 0.8890210 0.7480500 0.8926260 0.7150840   
%  0.0003640000 0.01900800 0.007678000 0.05214800 0.06676900
%  & \subfigure[]{\includegraphics[width=4cm]{figures/biomed/n5_50_samedice_worse_crop.png}} % count: 37
%30 0.3228370 2.314121 0.3088510 0.7239250 1.849457 0.3497480 0.9596710 0.9497610 0.9781700 0.8982020 0.7695680 0.9081760 0.8202250   
%   0.0002250000 0.01935000 0.007807000 0.05222200 0.06684600
  % \subfigure[]{\includegraphics[height=3cm]{figures/biomed/n5_50_samedice_good_crop_labeled.png}} % count: 30, dice6: .820
\end{tabular}
\caption{Classification of a slice of d-PET data using two different parameter configurations. \label{fig:samedice}
The classes are 1: background (BG), 2: skull (SK), 3: grey matter (GM), 4: white matter (WM), 5: cerebellum (CM), and 6: putamen (PN). 
%Despite the noticeably different shape of the putamen, the Dice coefficients indicate the same quality. 
%\sbnote{Remove numbering, if it does not appear in \SC{sec:biomedres}.}
}
\end{figure}

\subsection{Configuring a bio-medical image segmentation algorithm\label{sec:biomed} }
% High-level objective: Improve diagnostic means for higher quality health care.
%\sbnote{Get this description checked by Ahmed.} 
Saad et al.~\cite{Saad:2008:kinetic} use a kinetic model to devise a multi-class, seed-initialized, iterative segmentation algorithm for molecular image quantification. 
% Briefly explain the role of the kinetic modelling error and how the dice coefficients represent distance from training data.
Due to the low signal-to-noise ratio and partial volume effect present in dynamic-positron emission tomography (d-PET) data, their segmentation method has to incorporate prior knowledge.
In this noisy setting, the segmentation of a basic random walker~\cite{Grady:2006:RW} would just result in Voronoi regions around the seed points. 
A new extension by Saad et al. makes this method usable for noisy data by adding energy terms that account for desirable criteria, such as data fidelity, shape prior, intensity prior, and regularization. In order to attain the superior segmentation quality of the algorithm, a proper choice of weights for the energy mixture is crucial.

%- RWSL-KM overcomes this problem obtaining a significantly better segmentation for PN region; However, misclassifies a band around CM as PN. This maybe avoided with an improved parameter optimization
%- Dice coefficient does not take uncertainty (e.g. entropy of class membership probability vector) into account

%% What is a good segmentation? The dice coefficient measures similarity to training data.
%% Explain numerical objectives (dice coefficient and kinetic modelling error) and show optimality plateau for relevant dice.

To facilitate this choice of weight parameters, their code also provides numerical performance measures that assess the quality of each class. 
One such measure is the Dice coefficient~\cite{Dice:1945:dist}, which gives a ratio of overlap with labelled training data. A second measure expresses an error of the quality of the kinetic modelling. 
Overall, the algorithm is influenced by $8$ factors or parameters. Ten response variables provide the $2$ quality measures per class, disregarding background.

Theoretically, the parameter calibration could proceed by numerical optimization of the performance with respect to the weights of the energy terms. 
However, for instance the Dice coefficients that indicate agreement of the segmented shape with given training data for putamen, using the two configurations of \FG{fig:samedice}(c) and (d), are both above the 90\textsuperscript{th} percentile of the sampled configurations and less than $0.003$ standard deviations apart.
% In other numbers, they were 3rd and 4th best out of 50 with mean=$0.481$ and stdev=$0.281$.
Numerically, this means that both segmentations are of the same, near optimal quality. Yet by visual inspection, it is possible to tell that the putamen (PN) shape in (d) is favourable over the one obtained in (c).
Hence, guidance of a human domain expert is desirable to visually sort among several candidate solutions in order to find an improved segmentation, which is hard or impossible to choose automatically. % \R{3b}.
An interactive workflow that facilitates such a procedure is subject of \SC{sec:biomedres}.

% Why is user intervention required?
% 1. performance measures (derived response variables) help to automate the search for good parameters. They can be used to express necessary conditions, but may not be sufficient as shown in the example figure.
% 2. robustness of a parameter choice can be judged from the shape of the response surface.
% 3. we have multiple responses with different dynamic ranges, whose mixture also needs to be chosen.

% Further objectives: more dice and kinetic error measure.
% Goal for algorithm development: determine the importance of different energy terms for the quality of the segmentation result. 
% Sensitivity analysis or factor screening.

%Further, the resulting target function may be multi-modal with several different parameter sets leading to numerically acceptable fit with given training data.
%, while resulting segmented images still vary noticeably. %substantially. 

\begin{comment}
requirements:
- multiple criteria need to be traded off \R{4-6},
- multiple factors have to be assessed for their usefulness to steer towards more desirable segmentation results \R7,
- robust quality shows as large optimality plateaus, which amounts to visual sensitivity analysis \R7.

previous workflows only allow for random sampling with manual screening.
An overview of the computed data would be helpful to guide the search \R{3a}.
experiments are set up manually, using nested for-loops.
\end{comment}

% ----------------------------------------------------------------------------
\subsection{Engineering: Fuel cells \label{sec:cstack} }
A fuel cell takes hydrogen and oxygen as gaseous input and converts them into water and heat, while generating an electric current. 
Affordable, high-performance fuel cells have the potential to enable more environmentally friendly means of transport by greatly reducing 
%\ce{CO2} 
$\mathrm{CO_2}$
emissions.
To manufacture a prototypical cell stack costs tens of thousands of dollars. Hence, a reliable synthetic model can greatly bring down the price of finding an optimal configuration for production.

The example investigated here is a simulation of a fuel cell stack developed by Chang et al.~\cite{Chang:2007:cellstack}. Their stack model is a system of coupled one-dimensional PDEs describing the individual cells in the stack. It can be adjusted with about $100$ parameters, where suitable choices of values are known for most of these parameters from fitting to available measurements. Computing a simulation run outputs $43$ different plots that show how certain physical quantities, such as current density, temperature, or relative humidity vary across the geometry of the cell stack.
% Constructing a single fuel cell stack is expensive. 
The computer model can be rerun for different configurations and, thus, allows for much broader exploration of design options than real prototyping.
In particular, engineers are interested in varying different groups of parameters that represent the geometry of the assembly (size and number of cells in a stack), material properties (permeability), or running conditions (temperature, pressure, concentration) to study failure mechanisms and optimize performance. 
Experiments demonstrating the use of the suggested interactions are given in \SC{sec:cstackres}.

\subsection{Problem structure \label{sec:problem} }

All previous use cases are motivated by questions about a real-worl system. 
In each setting, domain knowledge about the problem has been expressed in form of a computational model and the parameter space of the model is explored in order to relate observations about its properties to the corresponding real setting.
The studies share a set of requirements as summarized in \FG{tab:requirements}.

% Requirements
\begin{figure}[ht] %% not using a table here, because its caption takes up an awful lot of space.
\parao{\hrulefill}
\begin{tight_itemize}
\item[\R1:] integrate with existing practices and code
\item[\R2:] specify parameter region of interest (ROI)
\item[\R{2a}:] sample ROI and compute data set
\item[\R3:] browse data providing overview (\R{3a}) and detail (\R{3b})
\item[\R4:] construct feature variables (assign manually or compute)
\item[\R{4a}:] combine features to derive a distance metric
\item[\R{5}:] identify region(s) of similar outcome in parameter space
\item[\R6:] find optimal point for a particular variable or user notion
\item[\R7:] analyse sensitivity of feature values to change in input
\item[\R8:] save state of the project for later reproduction
\end{tight_itemize}
\parao{\hrulefill}
\caption{
%Combined requirements from three use cases. 
Summary of the requirement analysis.
\label{tab:requirements} } 
\end{figure}

\para{Biological aggregation patterns} 
%Exploring the solution space of the model is important for verification or to study possible causes for certain emerging patterns. 
During our interviews, it became clear that it is important for these target users to inspect the behaviour of an existing \sname{MATLAB} implementation for their PDE system (\R1). 
This allows to invoke the simulation for different combinations of parameter values. %, where each computation takes about 5 to 30 minutes, depending on the resolution of the simulated pattern. 
Since multiple different parameter combinations have to be explored, it is necessary to narrow the computations down to suitable regions in parameter space (\R2) and to assist the choice of sample points in these regions (\R{2a}).
Visual judgement of the computed solutions (\R3) is one method to enable a qualitative distinction among different patterns of movement (\R{3b}) and to determine which different sets of parameter choices produce a given behaviour (\R{3a} and \R5).
The growth rate of a linear perturbation can be included as a feature variable (\R4). 
%Calling this function for each sample point constructs a dependent variable that indicates the stability type.
Due to the size of the space of possible solutions, computational help to generate an overview of all possible behaviours would be desirable.
%\R{4a}
To ensure findings are reproducible, it should be possible to save the state of the project (\R8).

\para{Bio-medical imaging} 
This use case is different from the others in that sampling and computation are not done by directly interfacing with the code, but rather are done offline to produce a data set for inspection. In this setting, assistance in choosing a suitable parameter region to sample is again an important task, where an interactive visual approach can be helpful (\R2).
Also, dependent feature variables (objective measures) are already constructed for segmentation quality (Dice coefficients) and kinetic modelling error. 
%
% The user has to choose a proper combination to optimize and find optimal plateaus.
Requirements \R{1-3} apply here, except for the sample creation \R{2a}. 
The comlexity of the segmentation problem can only be captured by multiple performance measures.
The main goal is to find a robust parameter configuration that leads to good performance and is robust under a number of varying factors. In particular, performance should be invariant for different noise levels or patient scans.
One step towards that goal is to produce a weighted sum of performance measures (\R4). However, automatic optimization (\R6) is challenging with multiple competing quality measures. 
%The choice of mixture can be guided by visual judgement of the quality of the produced image segmentation.
%A satisfactorily configured performance or quality measure may then be optimized  automatically (\R{6}). 
For the algorithm to work robustly under different factors, it is important that the segmentation quality does not decay too quickly for slight changes to the chosen input parameter configuration (\R7).
To enable the user to assess robustness of a performance optimum, it is helpful to identify the region of parameter configurations that lead to 'good' segmentation results (\R{3a+5}), in order to make a robust choice within it.

\para{Fuel cell stack design} 
This case of simulation model inspection invokes basic requirements \R{1-3}. The goal of constructing a high-performance cell stack is akin to \R4 and \R6. Before that, however, the need to have a reliable and trusted model requires to 
identify parameter region boundaries that indicate transitions in stack behaviour. 
Such a decomposition into distinct parameter sets (\R5) can greatly support reasoning about plausibility of the model.

% --------------------------------------------------------------------------------------
\begin{figure}[ht]
\centering
%\begin{tabular}{lcr}
%\subfigure[]{\includegraphics[width=4.5cm]{pics/engine-metam-green.png}} &
\includegraphics[width=8cm]{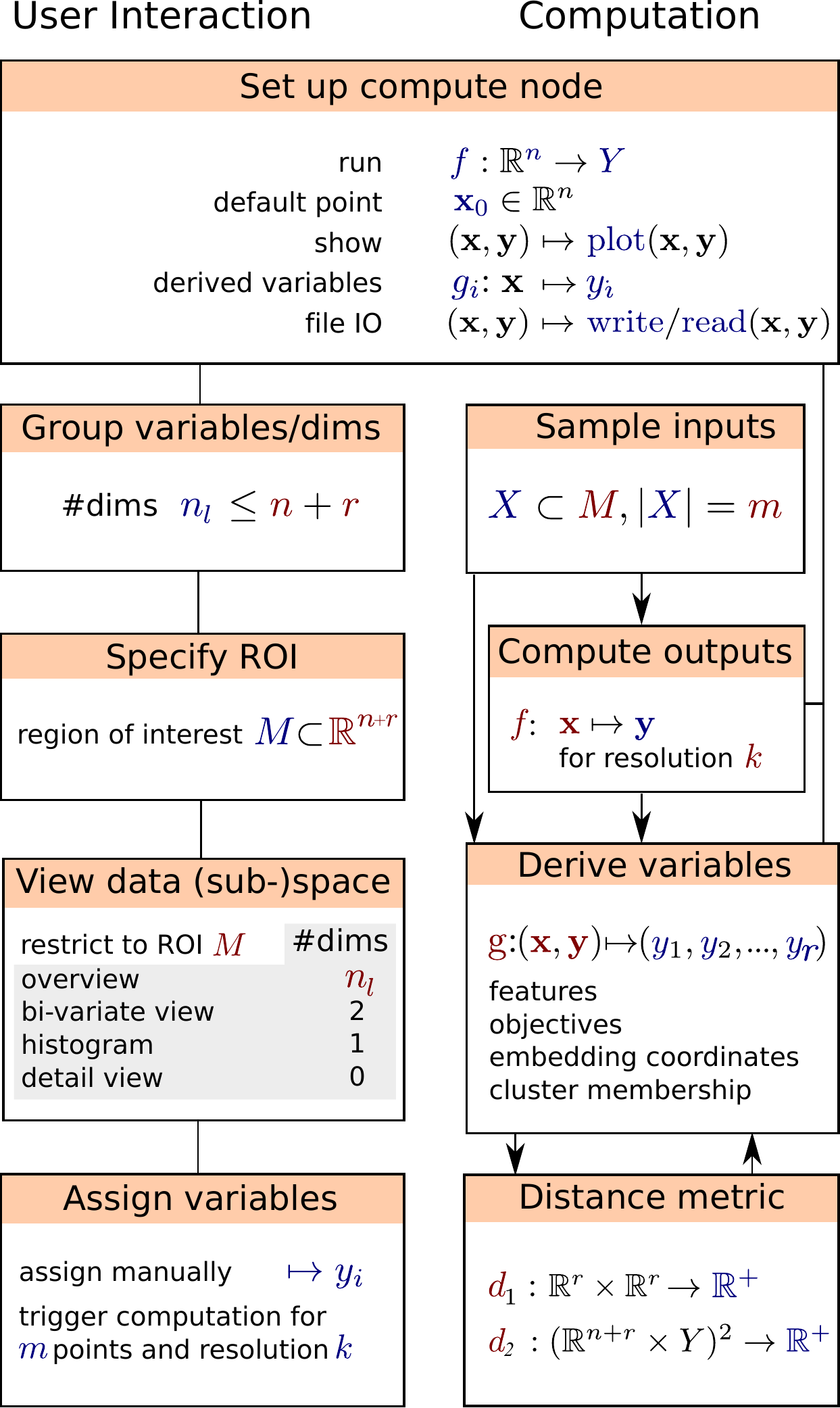}
%\includegraphics[width=17cm]{figures/conceptual-overview.png}
%c)\includegraphics[width=3cm]{pics/pal-col-gb.png} \\
%\end{tabular}
\caption{Abstraction of data, interaction, and computational components. Lines indicate shared data among processing steps and arrows prescribe an order of execution. On a more detailed level, 
{\color{BrickRed} Red} is required input and {\color{RoyalBlue} blue} denotes information that is available after a processing step. \label{fig:conview}}
\end{figure}
% --------------------------------------------------------------------------------------
\subsection{Data abstraction \label{sec:fun} \label{sec:abstraction} }

The conceptual organization of the required tasks is shown in \FG{fig:conview}. 
It separately considers user interaction and computational pipeline, where
all modules operate on the same data and share one flow of control. 
Integration with a computer simulation is possible using the compute node abstraction discussed in \SC{sec:compnode}.

In order to accommodate simulations with a large number of variables $n+r$, a first step of the interaction allows the user to divide them into groups of smaller size $n_l$, e.g. separating input and output, or indicating other semantic information inherent to the simulation.
The specification of a region of interest defines areas in the input space to sample and has further applications as detailed in \SC{sec:roi}.

%% begin of previous version
In order to capture relationships among variables one can combine their domains using a Cartesian product 
and express a relation as a subset of this combined tuple \defn{data space}.
For a functional computer model $f(\mb{x})\mapsto\mb{y}$ one can further distinguish between input \defn{variables} $(x_1,x_2,\dotsc,x_n)=\mb{x}\in\RR^n$ and output \defn{image} $\mb{y}\in Y$. While meant in a mathematical sense, $\mb{y}$ can represent an actual picture or a disk image that captures the result of the computation.
By application of another function it can be mapped into a Euclidean space $g: \mb{y}\mapsto(y_1,y_2,\dotsc,y_r)\in \RR^r$ of \defn{derived variables}. The $x_i$ and $y_i$ are also referred to as factors and responses,  respectively~\cite{Wong:1994:mdmv}. %% The terminology in this reference is somewhat outdated.
The inputs are often considered to be independent variables. However, this is not true in general, since the presence of constraints may introduce dependencies among the $x_i$.

Each \defn{configuration} or \defn{data point} $(\mb{x}_k,g(\mb{y}_k))$ represents parameter input and output for a run of the simulation with all $m$ points combined forming the rows of a relational data table. 
The columns of this table represent the variables, which are synonymously refered to as \defn{dimensions}. 
The variables that constitute input to the computation code are also referred to as \defn{parameters}.

While the function $f$ is itself deterministic, uncertainties in the system can be modelled by providing additional environmental variables $x_i$ that are distributed according to some probability measure~\cite[p. 121]{Santner:2003:DACE}.
%
% discretization
We assume that code to compute $f$ is available as a black box that can be invoked for a finite number of points $X=\{\mb{x}_k\}$. This set $X$ is referred to as a {\em design} or a \defn{sample}~\cite[p. 15]{Santner:2003:DACE}.
Together with the mapping $f$ this allows to compute the responses 
$\{g\circ f(\mb{x}_k)\}\subset\RR^r$. 

%Derived variables have a number of uses as features, objective or performance measures, dimensionally reduced embedding coordinates, or to indicate cluster membership. They may be assigned by hand or could be computed under potential use of a derived similarity metric.

Depending on what derived variable $y_i=g_i(\mb{y})$ is specified (\R4), its information may be interpreted as a feature, embedding coordinate, cluster membership label, likelihood for a model instance, distance from a template point, or objective measure --- to give a few practical examples. In each case it may be possible to compute values or to assign them manually, depending on whether a function definition or a user's concept is available.
Some processing steps require a notion of distance or similarity among points. It can be obtained as Euclidean distance $d_1$ over combined feature vectors in $\RR^r$. Beyond that, distance $d_2$ uses all information about each configuration point, including its parameter coordinates $\mb{x}\in\RR^n$ or a domain specific function operating on the disk images.

\ifthenelse {\boolean{optional_content}}{

\subsubsection{Relating to real world experience}

Inspection of this discrete distribution of points then may proceed using overview techniques for multi-dimensional multi-field data and should be complemented by detail views (\R{3b}) of individual data items. In all presented cases the detail views are domain specific. 
In the swarm model of \SC{sec:swarms} for example, showing detail means to depict the spatio-temporal movement pattern generated for the parameter configuration of a particular data point of interest.

As the user inspects properties of the model it is important to keep relationships with experienced or recorded properties of the real system in mind to verify plausibility of possible conclusions. 
This is also the stage, where the model output may be enhanced with further dependent variables. These could either be expressed in form of computable features or be manually chosen values, for instance allowing to apply tags or annotations to selected points. 

Up to this point the described operations are somewhat standard to multi-dimensional data exploration, except for the fact that the user also participates in generation of the data set.
The distinction that has been made in the introduction between optimization and discovery of distinct system states comes into play in the next step.
If the derived variables have been constructed to represent characteristic properties that help to tell distinct simulation outcomes apart, they can be used for distance computations among pairs of points.

Equipped with a meaningful distance measure, it is possible to assist the user in the discovery of distinct regions by grouping similar points. The clusters obtained in this manner may then be visualized with respect to the independent variables of the model producing a map of the input parameter space that separates regions of distinct model behaviour from each other.

The speed of such a transition between solution regions would be subject of a sensitivity analysis. 
While not directly part of such a visualization, it is possible to infer some sensitivity information from the extents of a cluster of similar points, where clusters that have a small extent in parameter space are more sensitive to changes.
This result is different from the end point of a performance optimization that seeks to find a particular solution that maximizes a number of desirable properties. Instead, the proposed parameter space segmentation represents a summary of properties of ensembles of computed runs. As such, it gives an overview of the capabilities of the model and may lead to a more comprehensive understanding of the real world system. 

%\ifthenelse {\boolean{optional_content}}{
%%%%%%%%%%%% report findings %%%%%%%%%%%%%%%%%%%%%%
\subsubsection{Report}
The implications of the so obtained findings should be reproducibly described in a report, concluding answers to the research questions from the outset of the study. This broad exploration process may also lead to serendipitous discoveries that should be noted as well. Quite commonly, the result of the investigation poses new derived questions that lead to further studies.
%
%To maintain or improve quality of the scientific discourse, it is important to identify the true source of the contributions, which should include credit for both, good questions and good answers. This is the basis for a sustainable reputation and reward system for research agents and their organizational structures.

}{} %end of optional_content

\begin{comment}
- to learn about cause/effect relationships G. Galilei (or F. Bacon?) suggested to produce tabulated recordings of multiple variables and to then compare how the values in the columns are changing with respect to each other 
- the general operation that is done here is a comparison of distributions, e.g. empirical ones made from measured data and structural ones, e.g. hypothesized distributions.
- to compare a pair of numerical variables, scatterplots have been the method of choice (cite some stats book, e.g. Santner or Global sens an.)
- effective methods to do this when the number of variables exceeds the dimensionality of the screen are still considered an open problem, current state of the art considers scatter plot matrices, parallel co-ordinates, or dimension reduction techniques.
- data acquisition
Since the sampling of the input and the computation of the output variables is controlled by the system, it is possible to include knowledge about adjacency information among the sample points to improve the rendition of the resulting continuous density distributions for the dependent variables, by adapting the cell projection algorithm of Shirley and Tuchman~\cite{Shirley:1990:PT}, as published by Bachtaler and Weiskopf~\cite{Bachtaler:2008:CS}.
\end{comment}

%------------------------------------------------------------------------------
\section{Background \label{sec:relwork}}

The following review will start with related systems that address certain requirements. 
Computational steering and experimental design will receive particular attention in this context. 
Methods specific to particular design aspects of \sname{paraglide} will be discussed in the respective sections of \SC{sec:design}.

\subsection{Interactive environments for parameter adjustment in computer experiments}

Computational steering considers adjustment of parameters during execution of time-dependent simulations~\cite{Mulder:1999:compsteer}. 
% Other system design requirement is seemless integration of the steering system with the domain specific code to be inspected.
Since our users do not modify parameters during the run of a simulation, our problem setting is different from classical steering in that we do not need to handle live updates of variables that are shared between simultaneously running modules. 
However, there is enough similarity to benefit from a comparison. % of design decisions.
% shared requirements: modular separation into parameter control, and domain specific computation code

%In the requirement analysis for 
An evaluation of their computational steering environment (CSE) by Wijk et al.~\cite{Wijk:1997:CSE},
%stress integration with existing systems and modular design as important consequences.
%
%Their system is built around a central data manager that stores variables and signals update events that satellites can subscribe to in order to re-compute their  derived variables.
%
%An evaluation of the system in 
recognizes major uses for debugging, presentation, and assistance in technical discussions that progress faster when "What if?" questions can be answered immediately. 
A follow-up survey by Mulder et al.~\cite{Mulder:1999:compsteer} identified further uses for model exploration, algorithm experimentation, and performance optimization. 
%Also, their mathematical models were found to be more useful to outsiders when represented as parameterized graphics objects, instead of simply vectors of values.
%
While these systems inspire numerous design decisions, the specific requirements for efficient regional sampling and an easy integration of end-user codes for simulation and derived variable computation are either not fulfilled or could be improved.

%The SCIRun system of Parker et al.~\cite{Parker:1995:SCIRun} 
%%present SCIRun, an integrated problem solving environment for computational steering.
%is based on a data-flow model. 
%% that entails a module (algorithm or operation), port (connecting point to route data), data type (scalar field, matrix), connection (one output to multiple input).
%Adding a new module involves writing C++ code, which for some MATLAB-trained end-users is a non-trivial hurdle to take before they can benefit from the framework. 

%Matkovi{\'c} et al.~\cite{Matkovic:2008:steer} present a system for visual steering to assist the generation of experiments in a multi-dimensional parameter space. A simplified model is produced by manually selecting a few relevant parameters, for which the specification of intervals of interest can be iteratively refined in further simulation runs. Linked views are shown to be useful to identify regions in the input parameter domain that cause certain behaviour in the response.
%The multi-objective optimization of their system also applies to some of our use cases. However, their system works on a pre-computed, static data set, which is too restrictive for most of our use cases. 
% of \SC{sec:requirements}.

Berger et al.~\cite{Berger:2011:mvpred} discuss a system to visualize engineering and design options in the neighbourhood around an optimal configuration of a computer simulation. Based on a continuous function abstraction they provide a local analysis method that benefits domain experts. In order to not get stuck in local maxima, optimization methods usually benefit from an additional global perspective on the problem domain. The qualitative decomposition pointed out in the cases of \SC{sec:requirements} and pursued in the following provides such a complementary view.

%We are not steering a time varying process, instead a static sample of points is produced for offline computation and interactive visual inspection.

%There is a recent rise of attention to the subject in the visualization community. 
%Matkovic 2005, 
%my thesis proposal 2006, 
%Vis session 2010 (FluidExplorer, Worldlines, Dimstiller~\cite{Ingram:2010:dimstiller}), 
The challenge of devising a user interaction for sample construction has recently been taken on in the \sname{Paranorama} system of Pretorius et al.~\cite{Pretorius:2011:paraspace}. 
Their users can specify different ranges of interest for individual variables along with the number of requested distinct values per range. The sample points are then constructed via a Cartesian product of the value sets. Integrating this method into an image segmentation system, received positive feedback from users. 
However, combining many value ranges with this method may result in large sampling costs%
\ifthenelse {\boolean{optional_content}}{
, as elaborated below in \SC{sec:dimfx}. %in the following section.
}
{. 
Beyond numerical arguments, also screen space real estate is used up more quickly when viewing data sets with large numbers of variables, and a significantly increased cognitive cost arises when investigating and interpreting the effects of many factors on possibly multiple responses.
%CONCLUSION:
Due to their significant impact on sampling and processing costs, \SC{sec:roi}
will give careful consideration to the number of involved variables and the volume of the region of interest.
} %end of optional_content

%Part of our work identified continuous visualizations \ednote{TM: it is not clear what you mean here with 'cont vis'?! I don't think it is even important to make this distinction here?! and again - this has nothing to do with 'computational steering'} and uncertainty guided sample refinement as fruitful research directions, which are elaborated in the adjacent work of Torsney-Weir et al.~\cite{Torsney-Weir:2011:tuner}. Their system presents a workflow to simultaneously optimize two performance measures. The approach presented below in \SC{sec:biomedres} and \SC{sec:cstackres}, 
%is an alternative to their method and allows to include a larger number of objectives. \ednote{TM: Perhaps you can move this paragraph down further, when you talk about this used case. You mentioned this paper here already, no need to do it twice.}

\subsection{Experimental Design \label{sec:rwsampling} \label{sec:gendat} }
% \subsection{Sample Construction \label{sec:rwsampling} \label{sec:gendat} }

%\sbnote{Some content of this section could be shuffled with \SC{sec:gendat}. Overall, it replaces a separate sampling discussion in this paper.}

\begin{comment}
:structure: 
- use: generate data set that allows to discuss the driving questions
- taxonomy (relationship to ROI): approximating one distribution with another
  - quality criteria: uniformity (integration error, reconstruction, geometry, regularity, periodicity, ease of inverting interpolation matrix, nested)
  - categorize existing methods: (non-)adaptive, initialization
  - fitting a model vs. model-free (uniform)
  - beyond uniformity: non-uniform density, non-euclidean domain
- how is it used in PG?, we just use boxes and specify additional parameters (p,r)
\end{comment}

The task of generating a data set for a function $f$ has been abstracted in \SC{sec:fun} as designing a sequence of points $\{\mb{x}_k\}=X \subset \RR^n$ that discretize its domain. 
%%% Indicating in \SC{sec:gendat} that costs are rising quickly as more variables have to be sampled. Hence, there is a whole research area of experimental design devoted entirely to this issue.
%Instead of descending into describing various types of sampling methods, let us consider different purposes of sampling:
%
While there is a vast body of literature on the subject, the point of the following discussion is to provide a flavour of relevant issues and give entry points on how they are approached. %While the system design has to consider efficient sampling as an important issue, 
%we do recognize that end-users should be concerned as little as possible with details of the sampling method. Before giving an overview of related methods for multi-dimensional sampling, let us consider different purposes of sampling:
\begin{enumerate}
\item Exploratory designs: To facilitate a comprehensive analysis of the function $f$, a model of its full joint probability distribution $P\{X\le\mb{x},Y\le\mb{y}\}$ could be constructed \cite[p.13]{Bishop:2006:PRML}, addressing \R{2a}.
%%where $X$ is distributed in the domain of $f$ like the given data and $Y$ the dependent distribution of responses. 
To obtain such a complete model, a possible approach is to use space-filling designs.
% with a space-filling property that are also studied in the context of uniform quantization. 
%%for $(\mb{x},\mb{y})$, such as Latin hypercube sampling, low-discrepancy sequences, or point lattices.
\item Prediction-based designs: 
In order to compute a single aggregate statistic, such as the mean value of $f$, one can numerically perform {\em integration}, which amounts to the application of a linear functional. Further settings might involve the application of a family of such functionals, e.g. to {\em approximate function values} at new positions, which is also studied under the name of {\em reconstruction} or {\em interpolation}. 
% In these settings the size of a sample has to be chosen to sufficiently reduce uncertainty (size of the confidence interval) of a predictor. This can also happen in a number of sequential updates.
In all of these settings it is possible to {\em estimate an error} or infer a confidence interval, which may or may not take newly acquired data into account. 
\item Reconstruction model adjustment: To further adapt the reconstruction or regression model to field data, sample points are used to determine an appropriate model family (linear/non-linear), a suitably reduced dimensionality, or other regularization parameters. For instance, dimensional reduction and the choice of a correlation function for Gaussian Processes fall into this category.
\item Analyzing variability:
  {\em Uncertainty analysis} determines variability of a response based on the distributions given for the environmental variables, which provides confidence intervals for the responses that can be used to guide selection of relevant features~\cite{Francois:2007:highd}. 
  {\em Sensitivity analysis} extends this analysis to determine how output variability is affected by each of the input variables~\cite{Saltelli:2008:gsa} (\R7).
  Similar measures appear in the analysis of (backward) stability of numerical algorithms~\cite[p.~104]{Trefethen:1997:NLA}.
\item Optimization-based designs:  
  In this setting, only those parts of the domain of $f$ are relevant that are likely to contain an optimum.
  Starting from an initial design, it is possible to steer concentration of the sample density in subsequent updates~\cite{Marks:1997:galleries,Brochu:2007:apl,Jones:1998:ego} (\R6).
\end{enumerate}
These tasks are somewhat ordered by decreasing degree of comprehensiveness. For instance, 
%unless a restrictive family of of parametric probability distributions is chosen, 
in the general case (1) will require an exhaustive sampling where (5) -- after suitable initialization -- may focus further points on small, promising regions in parameter space.

%We identified user needs to 
%(\R{2a}) create a data set inside a region of interest, and
%to (\R7) determine sensitivity in a given region or around a certain point.

\ifthenelse {\boolean{optional_content}}{

\subsubsection{Effects of dimensionality \label{sec:dimfx} }
A basic challenge arises from the number of variables that complex models entail. 
To see this, consider a classical factorial design, e.g. implemented via nested for-loops that iterate each variable $x_i$ over a set of values $X_i$. This leads to an overall number of %$m=\abs{X}=\prod_{i=1}^n \abs{X_i}$ 
$\prod_{i=1}^n \abs{X_i}$ points. Even if only two distinct values per variable are desired, this design has $2^n$ points, growing exponentially in the number of variables.
%This simple example is one aspect of a broader numerical effect regarding the sampling costs that are exponential in the number of variables. %when a certain coverage or reconstruction accuracy is required~\cite{Bungartz:2004:SG}.

Another effect arises from considering the Newton-Cotes quadrature method for multi-variate integration. It uses Lagrange polynomials of degree $s<m$ to properly represent smooth integrands \cite[pp. 119]{Phillips:2003:polyapprox}. The resulting error term is proportional to $h^{s+1}f^{(s)}(\xi)$ 
%\ednote{TM sure this is $h^s$ and not $h^{s+1}$?} \sbnote{fixed.}
for some $\xi$ contained in the integration domain with sampling distance%
\footnote{One way to obtain the distance $h$ from the number $m$ of sample points in a multi-dimensional setting is as edge length of the fundamental parallelepiped of a Cartesian lattice scaled to have $m$ points in a unit volume. 
%With irregular uniform sampling the same holds for a point density of $1/m$.
}
$h=m^{-\frac{1}{n}}$.
%%\ednote{Can this density be argued for without assuming a Cartesian arrangement?} 
%
This leads to an integration error $\varepsilon$ of order 
$O(m^{-\frac{s}{n}})$, where the bounded $(s+1)$\textsuperscript{th}-derivative of $f$ disappears as a constant factor. One can turn this around to estimate the cost $m$ for a desired accuracy as $O(\varepsilon^{-\frac{n}{s}})$, which is pointed out by Bungartz and Griebel~\cite{Bungartz:2004:SG} as an illustration of two important effects: Firstly, there is an exponential growth in the need for sample points with increase in dimensionality $n$ termed the {\em curse of dimensionality}~\cite{Bellman:1961}. Secondly, additional assumptions on the smoothness of the integrand, such as bounded derivatives $f^{(s)}$, may reduce this need and can thus be considered an example of a {\em blessing of regularity}.

\para{The volume of Minkowski-norm hyper-spheres} 
%is an interesting term to be aware of when studying multi-dimensional sampling. 
% It for instance appears as a proportionality factor in the estimate given earlier for the count $N_\epsilon(M)$.
%
%For a given volume, the sphere has the smallest surface area and is contained within the smallest diameter.
%
%Also, in Euclidean space $\RR^n$ with the Lebesgue measure the sphere is the body with the smallest $t$-neighbourhood (Minkowski sum with a ball of radius $t$) and surface area, which is referred to as isoperimetric scaling inequality (\cite{Matousek:2002:DG}, pp. 333).
An $n$-dimensional $p$-norm sphere $S_n^p=\{\mb{x}\in\RR^n : x_1^p+x_2^p+\dotsb+
  x_n^p\le r^p\}$ of radius $r$ with $p\in[1,\infty]$ has a volume as derived
by Newman~\cite[p. 101]{Newman:1972:IM}
\begin{equation}
\vol(S_n^p) = 2^nr^n\frac{\Gamma(1+1/p)^n}{\Gamma(1+n/p)},
\label{eq:psphere}
\end{equation}
using the Gamma function $\Gamma$ as a continuous extension of the factorial, giving $\Gamma(n)=(n-1)!$ for integral $n$.
The graph in \FG{fig:psphere} shows an interesting behaviour for increasing $n$
\begin{figure}[htb]
\centering
%\begin{tabular}{lr}
%  \subfigure[]{\includegraphics[height=5cm]{figures/p-sphere-n20}}
%  &
%  \subfigure[]{
  \includegraphics[width=8cm]{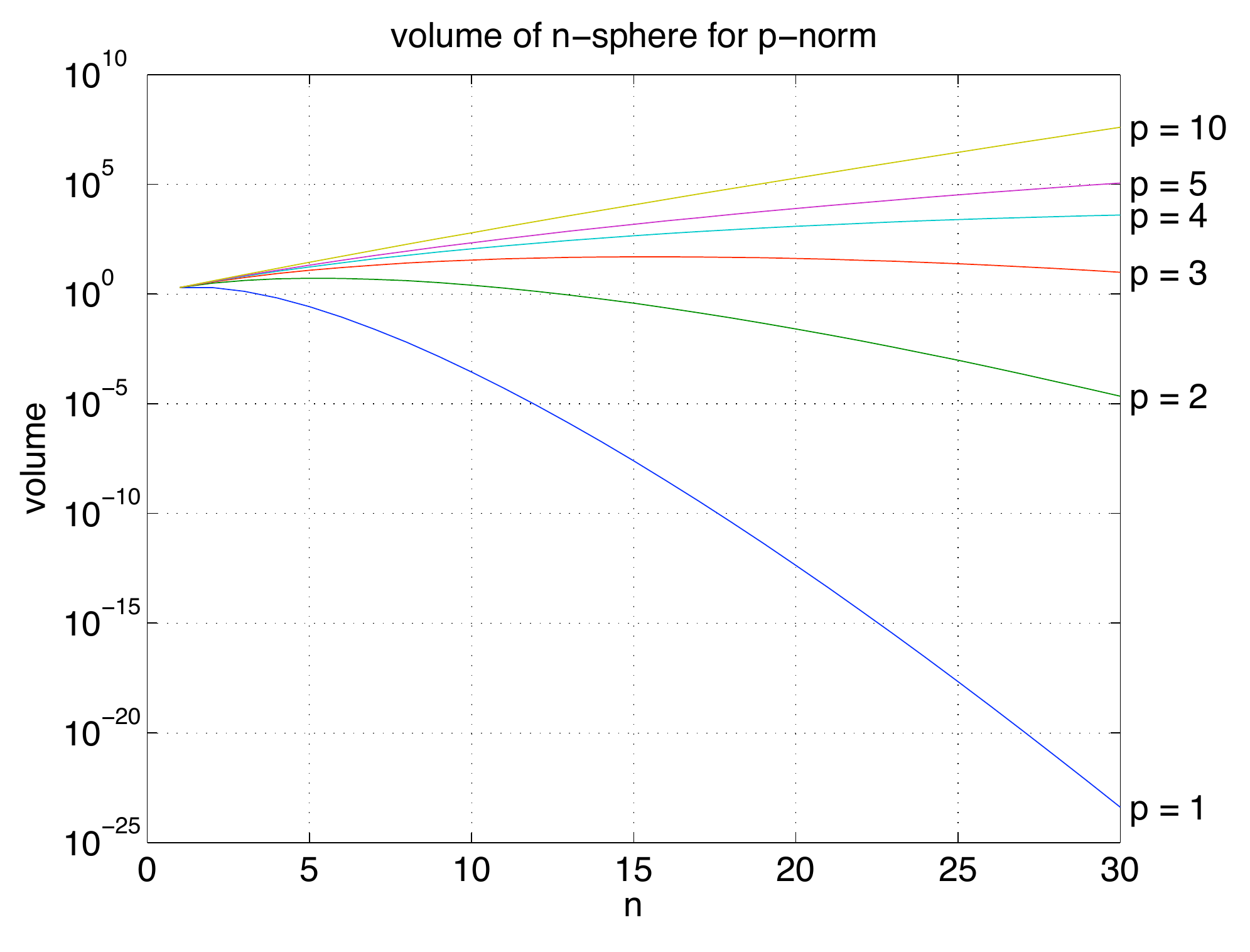}
%  }
%\end{tabular}
\caption{Semi-log plot of volumes of $n=1\dotsc30$ -- dimensional $p$-norm unit spheres.% in dimensions , approaching exponential growth for $p\rightarrow\infty$.
\label{fig:psphere}}
\end{figure}
with the $2$-norm sphere volume peaking at $n=5$ and volumes of all $p$-norm spheres ultimately converging to $0$ for all $n$ except in the case of the hypercube for $p=\infty$ with a volume of $2^n$.
It is somewhat misleading, however, to perform this interpretation along the $n$-axis, because a $3$-dimensional volume, for instance, contains infinitely many $2$-dimensional areas. However, comparisons along the vertical $p$-axis are fine and make a striking case for non-box shaped regions.

%Rearranging~\EQ{eq:psphere} one obtains the radius 
%\begin{equation}
%r_n=\frac{\Gamma(1+n/2)^{1/n}}{2\Gamma(3/2)}
%\label{eq:vol1radius}
%\end{equation}
%for an $n$-dimensional $2$-norm sphere of unit volume. 

% To show the effect of dimensionality vs. the blessing of regularity for Cartesian grids, one can consider the quadrature error of Newton-Cotes formulas~\cite[p.~5]{Lemieux:2009:MCQMC}.
Beyond numerical arguments, also screen space real estate is used up more quickly when viewing data sets with large numbers of variables, and a significantly increased cognitive cost arises when investigating and interpreting the effects of many factors on possibly multiple responses.
%CONCLUSION:
Due to their significant impact on sampling and processing costs, \SC{sec:roi}
will give careful consideration to the number of involved variables and the volume of the region of interest.

}{
} %end of optional_content

%\para{Non-adaptive sampling methods} 
%\bluenote{MS: No 3.3.1 when there is no 3.3.2}
The book by Lemieux~\cite{Lemieux:2009:MCQMC} gives an accessible overview on mostly non-adaptive (model-free) sampling methods that are for instance relevant to provide space-filling initializations for purposes 1, 2, or 4 in \SC{sec:gendat}. 
Another exposition by Santner et al.~\cite{Santner:2003:DACE} provides more background on model-adaptive sequential sampling, including an introduction to Gaussian process models (for purposes 2, 3, or 5), which are discussed more by Torsney-Weir et al.~\cite{Torsney-Weir:2011:tuner}. 

Of the above list, it will initially be aspect 1 that will matter in solving the problems laid out in \SC{sec:requirements}. As more insight on the model behaviour is gained and included in the analysis, the adaptive techniques of categories 2 and 5 will address sampling requirements more effectively.

\subsection{Parameter space partitioning}

The computational model analysis cases of \SC{sec:requirements} all benefit from an overview of regions of distinct system behaviour marked out in their input parameter space (\R5). There is no prior work in the 
academic visualization research
%area of computational visualization 
that provides such a representation. However, after considering different names for the method, such as parameter space segmentation, clustering, or partitioning, it is the latter term that relates us to two prior contributions from an old and a recent member of the sciences, namely physics and psychology.

Bhatt and Koechling~\cite{Bhatt:1995:paraparti} study the behaviour during impact of two solid bodies with finite friction and restitution, which results in a tangential sliding velocity that continuously changes direction after impact. 
The problem is characterized by $9$ parameters, three for the impulse direction of the impacting body and six for its rotational moment of inertia tensor.
The first step of their analysis determines a reduced set of three dimensionless parameters that completely define the tangential flow of sliding velocities.
An important observation is that the qualitative behaviour of the flow is characterized by $2$ or $4$ solution curves of invariant direction, as well as the critical points and sign changes of the velocity along these straight lines.
This results in $4$ main cases with up to $3$ sub-cases each. An implicit expression of the boundary between the cases is derived that is quartic in terms of the $3$ dimensionless parameters.
By fixing one parameter and showing slices through this boundary, the enclosed regions can be visually distinguished and are labelled with the different cases they represent. %see Fig. 5,7, and Table 1 
This provides a comprehensive overview of all possible sliding behaviour.
While providing a sophisticated algebraic analysis of a specific phenomenon, their discussion does not deal with numeric aspects involved of general computational models.

Pitt et al.~\cite{Pitt:2006:paraparti} also promote parameter space partitioning and give an example analysis of a model that predicts, whether visual stimuli are recognized as words or non-words. In their overview of analysis techniques, they distinguish two axes that separate quantitative from qualitative and local from global techniques. In this view, partitioning is a global, qualitative method, and sensitivity analysis a case of more local, quantitative inspection.

Their method proceeds from a notion of equivalence among model configurations and a set of valid seed configuration points. The regions around these points are sampled using a Metropolis-Hastings algorithm with uniform target distribution. Rejected points have fallen into non-equivalent, adjacent regions and are explored subsequently.

An important point made by Pitt et al. is to also address the need to improve the user's confidence in the plausibility of a given model. The studies they present are supported by showing the variety of qualitatively distinct model behaviour. However, a discussion of suitable analysis system design and considerations of required user interactions are not focus of their exposition.
%, rather than by an improved fit in a specific setting. 
%\ednote{TM - again, this is not a survey paper. How does their system solve or not solve our problems as detailed in section 2?}

%\subsection{Open research directions} 
\subsection{Features to contribute} 

Methods to visually inspect multi-variate point distributions (to address \R{3a}) are available in several of the frameworks listed earlier. However, the required capability to also generate data points (\R{2a}) or to add derived dimensions (\R{4a}) is missing from most systems that are mainly geared towards visualization of a static data set. Systems for experimental design, on the other hand, take care of the sampling requirements (\R{2a}), but often lack interactive, visual methods to solicit required user input (\R2). Computational steering systems combine sampling and visualization, but specifically focus on live-adjustments to parameters of a simulation that evolves over time. Their focus is often on some sort of interactive investigation, which
%optimization (\R6) 
could benefit from further support for broader state discovery (\R5).

% --------------------------------------------------------------------------------------
\section{Design of the \sname{\large Paraglide} system \label{sec:design} }

%\subsection{Conceptual structure}
%\ifthenelse {\boolean{optional_content}}{

%MS: Todo: Methodology intro
%We met with the participants during different phases of their work covering longitudinal time ranges of four years, two years, and five months with either monthly or weekly meetings, in the respective settings.
We will now discuss aspects of the design of \sname{Paraglide}, giving individual consideration to the graphical user interface (GUI), the software system, and choices of algorithms or methods for particular tasks.
\sname{Paraglide} was developed in a user-centered design process with five users, one or two from each domain. We met with our participants in person covering longitudinal time ranges of four years (fuel cell engineering), two years  (mathematical modelling), with monthly meetings, and five months (image segmentation) with weekly meetings. In these meetings we 
%(a) 
%\ednote{TM don't use letters for the listing here, since this gets confusing with the next section where they refer to something else. perhaps use roman numerals?} 
discussed design mockups and prototypes, 
%(b) 
observed our users working with \sname{Paraglide}, and 
%(c) 
gathered formative feedback in terms of usability and feature requests that we used to improve \sname{Paraglide}'s design. In addition, these meetings contributed to 
%(d) 
refining our understanding of user practices and design requirements (see \SC{sec:requirements}), as well as 
%(e) 
gathering summative feedback and anecdotal evidence (see \SC{sec:workflows}). 
%\bluenote{please check if this is what you did and extend if necessary}

%}{} %end of optional_content

%\sbnote{basic idea: nested for-loops $\rightarrow$ region + sampling method}

% --------------------------------------------------------------------------------------
\begin{figure*}[ht]
\centering
%\begin{tabular}{lcr}
%\subfigure[]{\includegraphics[width=4.5cm]{pics/engine-metam-green.png}} &
\includegraphics[width=17cm]{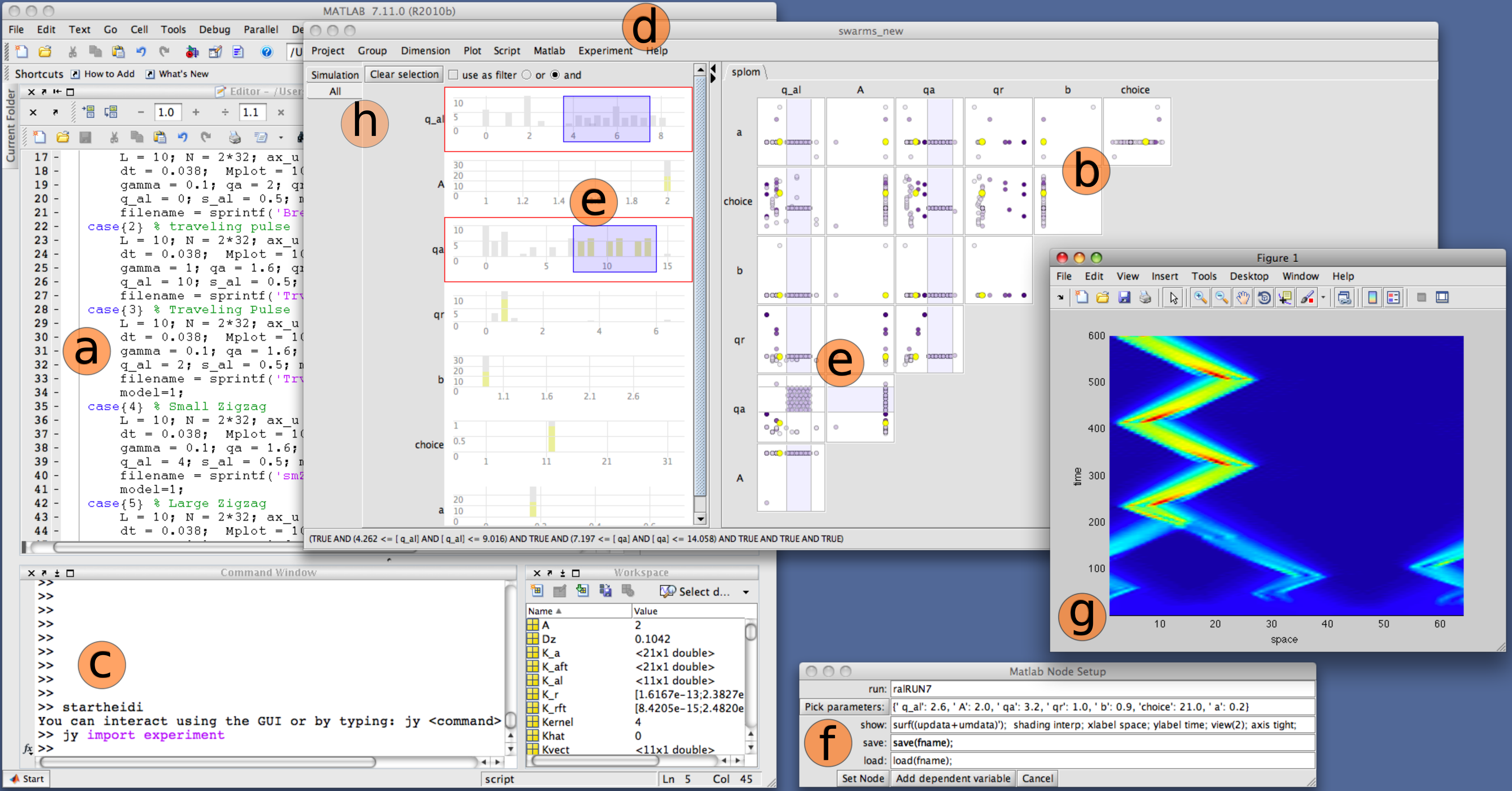}
%c)\includegraphics[width=3cm]{pics/pal-col-gb.png} \\
%\end{tabular}
\caption{Paraglide GUI running inside a \sname{MATLAB} session to investigate the animal movement model of \SC{sec:swarms}. \label{fig:screenshot} 
Initially, deliberately chosen parameter combinations are imported from a switch/case script (a) by sampling the case selection variable of that script and recording the variables it sets. An overview (b) of the data is given in form of a scatter plot matrix (SPloM) for a chosen dimension group (h).
Jython commands can be issued inside the command window (c) demonstrating the plug-in functionality of the system by manually importing the experiment module, which adds a new item to the menu bar (d). This allows to create a set of new sample points inside the region that is selected for parameters $q_a$ and $q_{al}$ (e). 
The configuration dialog for the \sname{MATLAB} compute node (f) sets up a {\em show} command that produces a detail view of the spatio-temporal pattern (1D+time) (g). For the configuration point highlighted in yellow in the SPloM, this results in a pattern of two groups that merge and then progress upwards in a 'zigzag' movement. 
}
\label{fig:gui}
\end{figure*}
% --------------------------------------------------------------------------------------

\subsection{System components \label{sec:gui} }
%\subsection{User interface components \label{sec:gui} }

%\ednote{This seems like it should be 4.2, not 4.1.1}
The snapshot of \FG{fig:gui} shows the \sname{paraglide} GUI and provides a brief overview of the main steps of the interaction.
%The menu bar on top of the main window allows to access basic functionality. Supplementary items added by imported modules may use the docstring of an associated Jython callback as tooltip help text.
%
In the left of the main window (\FG{fig:gui}d)
%(titled \sname{swarms\_new}) 
dimension group tabs are shown that can be used to switch between selected subsets of variables.
%dimensions \bluenote{MS: I thought we are talking about variables and parameters? Use consistent vocab!}.
Right next to it appears the view for an individual group of dimensions (h), 
which shows histograms indicating the distribution of values for the respective variables. If a group has more than $8$ dimensions, compact range selectors are shown instead of histograms. This frees up screen space and eliminates computational costs for keeping their information updated, e.g. when the data set or the filtered selection changes.
In \FG{fig:gui} the larger area in the right of the main window (d) provides a display of the data points (\R{2/a} + \R{3a/b}). 
In the example, a scatter plot matrix (SPloM) is shown (b) that arranges scatter plots on a grid, where each row or column is associated with one variable for the vertical or horizontal plot axes, respectively. 
Alternatively, it is possible to configure individual, enlarged scatter plots to inspect pairs of variables and show them in this area.
In the console in (c) one can enter commands for \sname{MATLAB}, \sname{Java}, or the \sname{Jython} interpeter that \sname{paraglide} is running.

\ifthenelse {\boolean{optional_content}}{

\subsubsection{Workflow integration via scripting}

Several requirements are shared among the use cases of \SC{sec:requirements}, but the tasks in the slightly differing application settings have to be addressed by different workflows. 
A basic requirement is to integrate with existing practices (\R1). Hence, \sname{paraglide} should support reuse of existing workflow ingredients, as well as flexible development of new constituents and their assembly. 
To construct a workflow for a task, a user should be aware of what the possible processing steps are and be able to specify how they fit together. 
%As alternative to visual programming, which is commonly used to this end\cite{RapidMiner,Dimstiller}), 
To facilitate this,
we simply provide a Jython scripting interface that is interpreted at runtime and operates in the same Java virtual machine (JVM) as a core library of Java classes. 
This allows more involved users to create plug-ins that may link with external libraries and other languages that are established in scientific computing, such as MATLAB, R, Python, or C/C++. 
Using the Jython \cname{import} command to load a module readily takes care of managing dependencies among plug-ins. 

Making the addition of plug-ins to the core system possible and easy enables the target user group to employ their own computation code. At the same time, it allows different research groups to independently pursue improvement of certain aspects of the framework and to share results. 
Several processing steps may be combined into simpler functions and can be made available for interactive use as mouse click callbacks or items on the menu.

% Examples: 
% cstack links in cell stack simulation, which comes out of a Java library that internally also uses \sname{LAPACK}.
% exeriment.createDataset() samples from a ComputeNode, which could invoke MATLAB, R, or Python code
% play.spiral() creates synthetic spiral data for some number of dimensions.
% Main script modules: jydi, experiment, matlab, embed, cluster, play, cstack, r
%To facilitate this form of manipulation, \sname{Paraglide} is implemented using Java and Jython scripting language, making it convenient to glue together any software components that provide a Java binding. 

%While developed as a stand-alone application, the \sname{Paraglide} system can also be invoked from within a MATLAB session, which enables bi-directional data and control flow. Since this is the current development environment for most of our users this conveniently integrates with their existing practices (\R1).

%\subsubsection{Data management, view, control, and state \label{sec:mvc} }

The core system is structured along the model--view--controller development pattern, which is partly inherited by using components of the \sname{prefuse} system~\cite{Heer:2006:DP}. 
Particular use is made of the ability to select points of a centrally maintained \sname{prefuse} table by evaluating a boolean expression on its row tuples with further details given in \SC{sec:roi}. 

%\sbnote{Stuff that we are happy to use: base table, expressions to select points, update signalling mechanism, control class for mouse events.}
% Activity manager is complicated to use and hence error prone.

\para{Reproducibility}
%The basic entity of a session is a \cname{Project}. 
To ensure that any possible findings made with the system are reproducible (\R8), the state of a project can be written to and instantiated from an \sname{XML} file.
It contains the filename for the base data table, a description of dimensions and their groupings, a selection region, current plot configurations, as well as miscellaneous property variables. 
A \cname{ProjectModule} abstraction enables plug-ins to add their own callbacks
for GUI updates or serialization, allowing to save them and instantiate from \sname{XML}. 
This includes a single line Jython initialization call, which allows to ship specific functionality along with a project description.
It is possible to use snippets of \sname{XML} inside a script to construct complex objects, such as SPloM views.
The \sname{XML} encoding also comes in handy, when just a part of the system state should be saved, e.g. to use the current selection region at a later time across similar projects. 
}{
\para{Workflow integration via scripting} 
Using the Jython \cname{import} command to load modules readily takes care of managing dependencies among plug-ins. 
It is possible to script workflows at runtime and add them to the menu. Dependent scripts can be stored and recovered along with an \sname{XML} description of the state of the current project. This also creates a separate folder that contains all cached disk images and other meta data.

\para{Data management, view, control, and state\label{sec:mvc}}
The core system is structured along the model--view--controller development pattern, which is partly inherited by using components of the \sname{prefuse} system~\cite{Heer:2006:DP}. 
Particular use is made of the ability to select points of a centrally maintained \sname{prefuse} table by evaluating a boolean expression on its row tuples with further details given in \SC{sec:roi}. 
} %end of optional_content

%\subsubsection{Compute node interface \label{sec:compnode} } % compute node abstraction
\para{Compute node interface\label{sec:compnode}} % compute node abstraction
To integrate domain specific computation code we use a \cname{ComputeNode} abstraction. It can return a list of parameter names with optional description text and set/get accessors. A node may have more specialized features that it can announce internally by returning a list of capability descriptors. The main ones are \cname{compute solution}, \cname{display plot}, \cname{file~IO}, 
and \cname{compute feature}, which in this order roughly correspond to the script lines that can be entered in the dialog of \FG{fig:compnode}.
Respectively, this means the node may be able to 
provide different detail plots for a solution,
it may compute solutions to a given configuration or 
derive named scalar or vector features, which are output quantities similar to plots. A node with \cname{file~IO} capability can store and retrieve cached solutions, such as \sname{MATLAB} data files.
%Similar to augmenting the computation for a data point with more derived variables, it is also possible to add more detail plots to the \cname{show} command. 
\begin{figure}[htb]
\centering
\includegraphics[width=8cm]{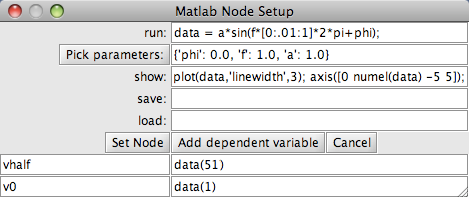}
\caption{Dialog to set up a \sname{MATLAB} compute node}
\label{fig:compnode}
\end{figure}
The \sname{MATLAB} node configuration dialog shown in \FG{fig:compnode} is a simplified interface for the \cname{ComputeNode} binding, where each of the edit boxes corresponds to one of the capabilities. 
In this particular example, the \cname{run} command creates a sine wave $v_t = a\sin( f t 2\pi+ \phi)$ for $101$ values of $t=0\dotsc1$,
parameterized by phase shift $\phi$, frequency $f$, and amplitude $a$, with default values 0, 1, and 1, respectively. The \cname{show} command displays the graph with axes of fixed height $\pm5$. Due to instant computation, \cname{file~IO} is disabled. 

The 'add dependent variable' button allows to enter a line of code, whose return value is assigned to a variable of chosen name. This implements the derived variables $y_i=g(\mb{y})$ described in \SC{sec:fun} and serves requirements \R{4/a}. 
If a scalar is returned, it can be shown in the SPloM view along with the input variables. 
It is also possible to return vector features that may not be shown directly, but can be used to compute similarity or distance matrices, or to determine adjacency information. Methods to derive further embedding coordinates from this information are discussed in \SC{sec:embed}.
The example in \FG{fig:compnode} 
picks out the deflection of the oscillation at time $t=0$ ($v_0$) and half way into the interval ($v_{\frac{1}{2}}$).
When inspecting these two 'features', it becomes apparent that $v_0$ depends on $a$ and $\phi$ only, where $v_{\frac{1}{2}}$ is influenced by all three parameters. 
In this simple setting, it is possible to make this observation by thinking about the equation given for $v$, as well as by studying the scatter plots of input/output variable combinations. For the latter method, however, we would first need to create a data set of tuples $(a,f,\phi,v_0,v_{\frac{1}{2}})$.
%, which is the topic of \SC{sec:sampling}.

With a readily configured connection to the simulation back-end, \sname{Paraglide} has a notion of the input parameter space $\RR^d$ as well as the output space $\RR^k$ of the simulation code. Initially, however, there may only be a single point of default values present, around which it is possible to expand the data set. 
To generate additional points it is possible to use some of the methods discussed in \SC{sec:rwsampling}. While \sname{paraglide} does not require to start with a given data set, the following discussion of system components will assume that an initial set of points already exists. 

%\bluenote {MS: Design and Implementation are still mixed. Personally, I do prefer a strict differentiation between these two. }
%\bluenote {MS: too long, too feature-driven instead of task/problem-driven. This is supposed to be a design study, right? So how does all the stuff address the problems identified earlier on... Maybe this comes later, but still I would love to see this section more in the light of problem-driven design decisions... }

%------------------------------------------------------------------------------
\subsection{Browsing computed data \label{sec:browsing}}

This stage provides the user with an overview of 
the data points as they distribute over input, output, and derived dimensions.
\begin{comment}
We will cover sub-topics in order of familiarity. For the workflows that will be discussed in \SC{sec:workflows}, however, this constitutes a somewhat inverse order of execution:
\begin{itemize}
\item viewing multi-variate multi-field data
\item grouping of variables/dimensions to reduce and organize visual and numerical complexity
\item constructing regions of interest
\item expanding a point
\end{itemize}
\end{comment}

%In order to employ standard viewing techniques for multi-dimensional data at least two points with distinct values in a number of dimensions in a number of dimensions.
%Views allow for detail inspection when the mouse hovers over a data point.

%\subsection{Multi-dimensional multi-field data visualization}
%\subsubsection{Visual mappings for multi-variate data \label{sec:vismap}} % Mapping data dimensions to screen space
% Mapping data dimensions to screen space
\subsubsection{Viewing multi-dimensional data spaces\label{sec:vismap}} 

To address requirements \R{3a/b},
we provide multiple simultaneous plots that give different views of the same data table and are linked to display a common focus point, highlight, or selection region. 
The subspaces that can be visualized that way range from multi- to 0-dimensional, 
to allow the user to relate overview of the whole distribution and detail plots that represent a single point.
We also provide techniques to view 2D subspaces, such as scatter plots that allow for pair-wise inspection of relationships among variables, and 1D projections of the marginal densities that can be shown in form of histograms.

There are many more possible techniques for multi-variate multi-field data visualization, such as
parallel coordinates, star plots, biplots, glyph-based visual encodings, or
scatter plots arranged in a matrix or table layout, 
with discussions of pros and cons provided in various surveys~\cite{Holbrey:2006:drvis,DeOliveira:2003:expomine}. 
%\ednote{TM: this was all review, nowhere in this section are you stating what exact plots you are providing!?}
Implementations of these techniques are either contained directly in \sname{paraglide} or are available via export to \sname{MATLAB}, \sname{R}, or \sname{protovis}.

%De Oliveira et al. relate multi-dimensional visual exploration and data mining, observing a need for methods that more tightly integrate visualization and analysis.

%Visual encoding should be clear. To help the user to interpret the spatial embedding one can provide axis labels, or possibly a coordinate grid. As a side effect, this way a  plot may also make sense without any points in it.

\subsubsection{Grouping dimensions}

As mentioned at the outset of this paper, the increased number of variables involved with modern simulation codes poses a challenge for their cognitive and numerical analysis. The visual complexity rises and makes data plots more difficult to interpret. 
The strategy of combining multiple views has its limits in screen space, as well as perceptual and cognitive handling. A possible remedy to this problem could be a) indirect visualization of fitted reduced dimensional models~\cite{Holbrey:2006:drvis}, or b) to divide the overall number of dimensions into groups for more focussed inspection.

%Holbrey~\cite{Holbrey:2006:drvis} indicates a trend towards indirect visualization that includes additional levels of processing, mostly involving fitting of alternative representations that are then visualized instead of the original data.
%In particular, past efforts from the statistics, data mining, and visualization communities are discussed and an overview of related data reduction and machine learning techniques is given.
%For that reason data analysis methods from statistics and machine learning have entered the field.

Grouping simplifies complexity and can be based on statistical or structural information.
%In \SC{sec:pointgrouping} we will discuss a user interactions to do this for points. 
Research on grouping of variables may consider dimension reduction and feature selection. 
To allow the user to express semantic information, we provide an interface to construct or modify a dimension group that simply consists of check boxes that indicate group membership for each variable.
During browsing, only dimensions of the currently selected group are shown, which reduces the required screen space. 
While automatic assistance in forming these groups is imaginable, our current approach of manual selection proved sufficient in all use cases.

\subsection{Specifying a region of interest \label{sec:roi} }

\begin{comment}
structure:
- usage for viewing and sampling and labelling and comparison
- abstraction: prob. dist.
- challenge: 
- related work
- interaction design
\end{comment}

% TASKS
In the requirement analysis of \SC{sec:requirements} it became apparent that 
the specification of a region of interest (ROI) has multiple uses in different sub-tasks:
\begin{itemize}
\item specify a domain or sub-regions for sampling to create or refine the data set,
%\item specify a bounded domain to generate a sample of points %construct a discrete sample of points inside the region
\item choose a viewport to focus the overview,
\item steer a cursor to set default values or invoke a detail view,
%\item a degenerate region, shrank to a point is a cursor, which may be used to specify defaults and focus for exploration
\item make a selection of points for subset processing, for instance to manually assign labels,
\item filter points to crop the viewed data range in order to deal with occlusion,
\item enable mouse manipulation of the region description,
\item export/import region descriptions to compare among different data sets.
%\item non GUI adjustments to the focus region, using a mixing board, can benefit tasks, where focus is required to assess visual quality of plots related to the selected points
%\item specifying a distribution by hand is investigated in the statistical learning community under the name of prior elicitation
\end{itemize}

% ABSTRACTION
\subsubsection{Representing a region \label{sec:roiabstract} }

%\sbnote{Link this more tightly with \SC{sec:abstraction} and \SC{sec:gendat}.}

Choosing a region of interest (ROI) and the construction of a set of sample points inside it, as described in \ref{sec:rwsampling}, are tightly related. 
The shape of the region and a locally varying level of detail amount to support and density of a probability distribution. 
%\ednote{TM - unclear, improper grammar} \sbnote{improved?}
% directional level of detail also requires a correlation function or tensor matrix
Seeing a finite set of sample points as discrete and a region description as continuous distribution, one can use distributional distance measures to assess how well one approximates the other.
%
%The region construction will be discussed separately in \SC{sec:roi}. In the following we will focus on generating the data.

\begin{comment}
Example for anisotropic tensors:
- electric permittivity or elastic stiffness of anisotropic media
\end{comment}

While conceptually a probability distribution sufficiently describes what is needed to capture about a region of interest, it may be difficult for the user to grasp or specify, especially in multi-dimensional domains.
A possible simplification is to omit the varying level of detail and to just consider uniform density. In this view the region is given by the support of the distribution and its volume corresponds to the inverse density.
Defining more complex regions than hyper-boxes in Euclidean spaces of possibly more than three dimensions, however, is a complex task for a human user. An algebraic way to express what is wanted, as pursued with the feature definition language (FDL) of Doleisch et al.~\cite{Doleisch:2003:FDL}, can provide complementary input that goes beyond the expressive power of current interaction widgets.
The \sname{XML} encoding of \sname{paraglide}'s system state includes a region description, which can be separately stored and imported.
%
%This is the direction pursued with the feature definition language (FDL) of Doleisch et al.~\cite{Doleisch:2003:FDL}.
%allows to specify hyper-boxes in a continuous parameter space.
%A 'feature' in their language is a selected region (specifically unions or intersections of ranges) along with a degree of interest, that may decay from $1$ to $0$ towards the boundaries of the region. Region bounds are defined per dimension in terms of threshold values (or percentage of the viewing range) and are then combined via OR/AND in a tree-structured language to form a multi-dimensional selection. Using bounds to specify regions rather than listing the selected points is called 'data set independent'. 
%\ednote{This is a section that could potentially be cut. You are touching on many things you don't do in the paper. Perhaps you should cut them out, and enhance the clarity as well.}

%\subsubsection{Goodness criteria and types of regions}
\subsubsection{Beyond the box --- filtering derived variables \label{sec:beyondbox} }

While this prior abstraction prepares much of what is needed in our application settings, the principal region template of a hyper-box might prove impractical in a higher-dimensional setting involving many variables.
% How dimensionality affects volume:
%\sbnote{If the sampling section came earlier, we could put this discussion there.}
The reason for this lies in the drastic way the volume of a hyper-box rises relative to an inscribed 2-norm sphere as their dimensionality increases.
This may not be much of a concern when selecting points from a given set. When generating samples, however, the costs are usually proportional to the volume of the requested region.
%However, as was discussed in \SC{sec:gendat} the volume of a region matters very much, \ednote{TM you don't discuss this anymore in section 3.2} when one seeks to represent it with a set of points. 
So, ideally one would like to keep it as small as possible.
% Enclose region of interest tightly, keep volume low to require less samples to fill.

The degenerate form of a region shrunk to a single point constitutes a {\em cursor}. 
It can be used as a focus point for detail inspection, as well as a method to choose a default anchor for the generation of new samples.
%, as will be described further in \SC{sec:sampling}.
%
Starting from this, one way to obtain smaller regions simply amounts to a change in perspective from defining a region that covers certain value ranges to constructing a meaningful neighbourhood around the focus point.
A common construct for such a point neighbourhood is a sphere bounded by some radius in a given metric.
This points to an alternative method of constructing custom regions by forming a hyper-box of ranges for chosen derived variables:
For polytopes these would be linear combinations of other variables that make up the plane equations for the bounding facets.
To obtain spheres in any metric, one could create a variable that measures the distance from a focus point. Bounding the maximum of this distance variable implicitly constructs a sphere on the dimensions that were involved in the distance computation. Boxes in this view are represented by the $\infty$-norm and a simple switch to the isotropic Euclidean $2$-norm sphere can significantly reduce the volume of interest. 
A data-adaptive example of dependent variables that facilitate, interactive multi-dimensional point selection is discussed in \SC{sec:embed} and applied in \SC{sec:biomedres}.

\subsubsection{User Interaction}

The input method that we use to specify a region of interest is similar to prior approaches for constructing hyper-boxes.
%
% There are several prior works on user-driven region construction.
%
% RELATED WORK
One of the first approaches is the HyperSlice system by van Wijk and van Liere~\cite{van:1993:hyperslice}. They steer a multi-dimensional focus point by constraining its coordinates using multiple clicks that locate its position in different 2D projections that are presented in form of a scatter plot matrix.
%
%They provide a visual interface for interactive brushing (selection) on linked views
% There is no automatic method of feature selection available in their system, as it could be provided by dimension reduction algorithms.
%that is similar to the interaction suggested by 
Martin and Ward~\cite{Martin:1995:hdbrush} extend the possible user interactions to form a hyper-box shaped brushing region beyond scatterplot matrices to include parallel co-ordinates, or glyph views.
% devise mouse interactions to specify a hyper-box shaped brushing region in different views, such as scatterplot matrices, parallel co-ordinates, or glyph views.
%
The prosection matrix by Tweedie and Spence~\cite{Tweedie:1998:prosection} also provides a similar form of control where the cursor point can be expanded into a hyper-box. The data inside the box is projected into different scatterplots. Collapsing an interval to a point changes the corresponding view from slab projection to slicing.

The previous interfaces map the data space to screen space using multiple axis aligned projections to $1$ or $2$-dimensional subspaces that map to range sliders or scatterplots. This requires the user to make a sequence of adjustments in order to specify a single cursor or region position. While this allows for a precise placement, the time required to make an adjustment grows at least linearly in the number of dimensions, while requiring the user to attend to multiple controls.

\subsubsection{Non-linear screen mappings \label{sec:embed} }

One way to reduce the complexity of multi-dimensional cursor or region control is to reduce the number of involved linked projections. 
This could be achieved by providing views that give more comprehensive information about the data distribution than 2D projections. 
In particular, there is a family of techniques for non-linear screen space embeddings that are designed to reveal most characteristics of the data distribution. Enabling the user to make selections in such a view may obviate the need to consider views from other angles.

In these methods, each experimental run is again represented as a point, where spatial proximity among points corresponds to similarity of two runs. Point placement with respect to the coordinate axes is typically hard to interpret.

%\ednote{Explain the problem first and make the explanations a bit easier to follow.}
Prior work into this direction constructs slider widgets for smooth $n$-D parameter space navigation, as developed by Bergner et al.~%in our previous work
~\cite{Bergner:2005:PSVR} and by Smith et al.~\cite{Smith:2007:navishape}. Both present a 2D embedding of sample nodes to obtain coordinates based on the screen distance between a movable slider and a set of nodes. Any curve the user describes by dragging the slider results in smoothly progressing weights that interpolate data at the nodes, which could result in different mixtures of light spectra or shape designs in the respective application setting.
%Check what they do for embedding, e.g. order in circle, pp. 1554
%
%\ednote{TM - here you are switching from a UI-driven dim-reduction to general dim-reduction, withouth mentioning dim-reduction?}

\para{Dimension reduction} 
Instead of arranging points in a circle or another prescribed shape, it is also possible to place them in a data adaptive way using dimension reduction techniques.
Kilian et al.~\cite{Kilian:2007:shapespace} use a distance preserving embedding of points to represent shape descriptors to control different designs of shapes. J{\"a}nicke et al.~\cite{Jaenicke:2008:brushing} lay out the minimum spanning tree of the data points for a multi-attribute point cloud, allowing the user to specify a region of interest in this embedding.
Extending this to larger sets of points, the \sname{glimmer} algorithm of Ingram et al.~\cite{Ingram:2009:glimmer} is able to produce distance preserving embeddings for thousands of nodes via stochastic force computation on the GPU. 
%\ednote{TM - So, glimmer is fast MDS. How can you mention glimmer without mentioning MDS and, in general dimreduction methods?}
%
Each of these methods require some notion of distance or adjacency among points, which is derived from dependent feature vectors that can be constructed 
through the interface described in \SC{sec:compnode}.

\para{Spectral embedding}
% Our method of using spectral embeddings is similar to asymmetric correspondence analysis or biplots.
To embed $m$ data points from an $n$-dimensional space to the $2$-D screen~\cite{Luxburg:2007:spectutor}, 
we start from a data matrix $X\in\RR^{m\times n}$. 
It can be turned into an affinity matrix $A=XX^T\in\RR^{m\times m}$, whose elements are normalized as 
%$(C)_{i,j} = \frac{(A)_{i,j}}{\sqrt{(A)_{i,i}(A)_{j,j}}}$ 
$(C)_{i,j} = (A)_{i,j}/\sqrt{(A)_{i,i}(A)_{j,j}}$ 
to yield a correlation matrix $C$. 
This implicitly scales each row-tuple in $X$ to lie on the surface of the $n$-dimensional $2$-norm unit sphere. Hence, the operation is referred to as {\em sphering} and the resulting elements of $C$ may be interpreted as cosine similarity or Pearson's correlation coefficient. 
The orthogonal eigen-decomposition of this positive semi-definite matrix $C=VDV^T$ gives unit eigenvectors $\mb{v}_i=(V)_{:,i}$ with decreasing eigenvalues $\lambda_i=(D)_{i,i}$. The $m$ components of $\lambda_2\mb{v}_2$ and $\lambda_3\mb{v}_3$ are providing the $x$- and $y$-coordinates for the spectral embedding of the $m$ points, respectively.

Many variations of this technique are possible. 
Firstly, instead of constructing affinity $A$ via dot products, it is possible to employ any other 
%positive semi-definite 
monotonously decreasing
kernel $(A)_{i,j}=\varphi(\mb{x}_i,\mb{x}_j)$, such as the Gaussian similarity kernel $\varphi(\mb{u},\mb{v})=\exp(-\norm{\mb{u}-\mb{v}}^2/(2\sigma^2))$.
Often, sphering is combined with a previous {\em centering}, where the mean $\boldsymbol{\mu}=X\cdot\one/n$ is subtracted from the data $X$.
Alternatively, if the rows of $X$ contain frequencies or counts, one can rescale their sums to $1$, which projects the data points onto the surface of the positive orthant of the $1$-norm sphere.

The described method is dual to its popular ancestor --- principle component analysis (PCA), where one instead begins with affinity $A=X^TX$ of centred data $X$. The above embedding algorithm then gives the $n$-dimensional principal components $\mb{v}_i$.
When the data is projected onto an axis of direction $\mb{v}$ the variance of the resulting coefficients is
$\sigma^2_{\mb{v}} = E_X[((\mb{x}-\boldsymbol{\mu})^T\mb{v})^2]
 = \mb{v}^TE_X[(\mb{x}-\boldsymbol{\mu})(\mb{x}-\boldsymbol{\mu})^T]\mb{v} 
 = \mb{v}^TC\mb{v}$.
%\begin{eqnarray}
%\sigma^2_{\mb{v}_i}&=&E_X[((\mb{x}-\boldsymbol{\mu})^T\mb{v}_i)^2] \\
% &=& \mb{v}_i^TE_X[(\mb{x}-\boldsymbol{\mu})(\mb{x}-\boldsymbol{\mu})^T]\mb{v}_i \nonumber \\
%  &=&\mb{v}_i^TC\mb{v}_i=\lambda_i. \nonumber
%%$=\mb{v}_i^TE_X[(\mb{x}-\boldsymbol{\mu})^T(\mb{x}-\boldsymbol{\mu})]\mb{v}_i=\mb{v}_i^T C \mb{v}_i$
%\end{eqnarray}
This shows that the correlation matrix can be used to compute variance of the data for arbitrary axes and explains the special role of its dominant eigenvectors $\sigma^2_{\mb{v}_i}=\mb{v}_i^TC\mb{v}_i=\lambda_i$ to give directions of maximum variance.
The different initializations of the data matrix lead to a family of techniques that include biplots, correspondence analysis, and \mbox{(kernel-)}PCA --- all of them may be efficiently computed via singular value decomposition (SVD)~\cite{Greenacre:1987:biplotca}.

\para{Grouping points} 
%Clusters of sample points are formed manually by outlining the boundary in a plot of choice.
%
Clusters of similar sample points can have arbitrary shapes. To not impose too strict assumptions on their distribution we opt for a manual method to assign cluster labels. For that, the user determines a plot where the clusters of interest are sufficiently separated. % from the remaining points.
After enclosing the points by drawing rectangular regions in the plot, it is possible to manually assign cluster labels or to directly work with the multi-dimensional region description. Automatic clustering algorithms could perform poorly in this setting, if assumptions about cluster shape, such as convexity, are not met.

%\ednote{TM - if it were that easy, than why not do automatic clustering? "where the cluster ... is clearly separated" is the actual problem.} 
%at early stages of model development it is often premature to impose strict assumptions on the shape of the clusters and enabling user intervention can provide the required flexibility.

%\sbnote{The section starts with the problem: reduce complexity, help with grouping points}

% ----------------------------------------------------------------------------------
\section{Validation in different use cases \label{sec:workflows} }

In this section, we discuss qualitative feedback and anecdotal evidence from user interviews and usage sessions during the later stages of our user-centered design process (see \SC{sec:design}).
The first stable and iteratively improved versions of \sname{Paraglide} were installed in the work environment of our three lead users.
%We iteratively provided our users with updated versions of \sname{Paraglide} when we added additional features or improved usability aspects. 
Overall, \sname{Paraglide} has been in use at several occasions over a period of $4$ years, accompanied by weekly meetings of our research focus group for a period of $2$ years, and problem-adaptive meeting frequencies in the domain settings.
% of $1.5$ years on a weekly basis, and we met with them every other week.

We present the summative findings in form of usage examples in which our users (a) were able to do something that they weren't able to do without \sname{Paraglide}, (b) could gather new insights into their model by using \sname{Paraglide}, or (c) felt to be able to conduct some task more efficient with \sname{Paraglide} than with traditional tools. 
\subsection{Movement patterns of biological aggregations \label{sec:swarmsres} }

\para{Semi-automatic screening for interesting solutions} 
%%\para{Adding a feature extractor} \\
%\sbnote{
%Problem: Support pattern search by screening for unstable solutions first, because those ones are hypothesized to produce interesting patterns. \\
%Method: a) perform linear stability analysis, b) implement stability type feature extractor as MATLAB function, c) add dependent variable to paraglide and color by it \\
%Added benefits: automatically identifies potentially interesting ranges for parameter choices}
In \SC{sec:swarms} a conjecture was pointed out that 
interesting spatial patterns only form with parameter configurations for which the PDE has unstable spatially homogenous steady states.
%This method perturbs a given steady state by adding spatially non-homogeneous terms (oscillations of different wave numbers). 
%\sbnote{Moved this here from 2.1}
%Substituting these perturbations into the PDE system allows to obtain their growth rate. Positive growth indicates an unstable solution and negative growth represents stability.
% denoted by $\sigma$.
%Linear stability analysis gives the stable and unstable wave numbers of the perturbations.
%
% No patterns can be observed for the stable wave numbers since the perturbations damp out. 
This means that linear stability analysis of \SC{sec:swarms} can be used to detect the potential for pattern formation.

In the course of this research project, the model developer
%analyst \ednote{I hate that word - what's an 'analyst'?} 
implemented a function to compute the type of (in-)stability for a given steady state. She had no problems to make the feature available in \sname{paraglide} within few minutes using the compute node interface of \FG{fig:compnode}. Colouring the data points by stability type then helped to focus the pattern search, because computing the feature based on just the input $\mb{x}$ takes about $5$ seconds, where a full population density would take $5$ minutes per configuration point. With this computationally cheap screening, it became possible to cover larger areas of the parameter domain before zoning in on sub-areas to compute more comprehensive output that includes spatio-temporal patterns. 

During one meeting a simple positivity test was implemented this way to answer, within five minutes, whether any solutions with negative densities were present --- providing a very efficient debugging aid.

%\para{Generating data} \\ % {Create new sample points}
\para{Discovering structure}  % {Create new sample points}
%\sbnote{
%Problem: To analyse multi-dimensional black-box function \\
%Method: a) construct ROI, b) produce sample point sequence, c) compute and cache, d) inspect
%} 
%\sbnote{This one could be moved into section 4.}
%\bluenote{Hmm,no rewrite I think this one is actually pretty interesting, but still the interesting things seem to hide somewhere ...
%Try to sum (for table): Novel insights into parameter space, feature: sampling, ... }
%
%\sbnote{Include discussion of scaling computational complexity via choice of resolution $k$ in \SC{sec:compnode}.}
To generate a set of sample points simply by specifying the containing region, the uniform sampling method, and the requested number of points,  was considered a very convenient way to generate data: {\em ``You don't need to worry about the coding, e.g. for loops, to set up region bounds, or choose sampling strategy.''} Aside from saving time, the interaction also puts the user's focus on core questions of choosing and combining value ranges.
%Our user found that \sname{paraglide} made it easier to define an experimental region with the ROI abstraction discussed in \SC{sec:roi}.
Within the selected range, coarse sampling to provide overview, followed by more focussed, finer sampling to acquire details, proved to be a good strategy --- the structure in \FG{fig:stabcorner} was found this way and inspired our user to further analytic investigation.
%
%To gain a better understanding of the stability region a finer resolution grid is placed over the sub-region, where the stability transition is expected to occur. 
%The corner in the boundary of this region was found with this method and inspired our user to further analytic investigation.
In particular, that an increase in repulsion leads to increased stability and less pattern potential corresponds well with biological experience.
\begin{figure}[htb]
\centering
%\begin{tabular}{lcr}
%\subfigure[]{\includegraphics[width=4.5cm]{pics/engine-metam-green.png}} &
\includegraphics[width=7cm]{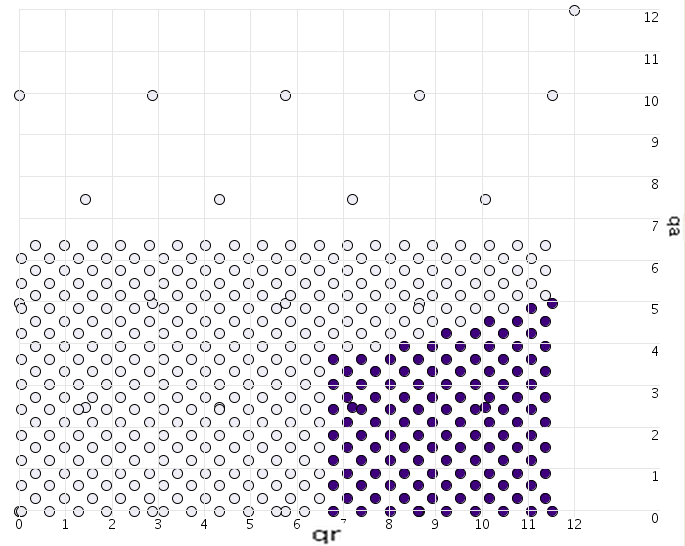}
%\includegraphics[width=17cm]{figures/conceptual-overview.png}
%c)\includegraphics[width=3cm]{pics/pal-col-gb.png} \\
%\end{tabular}
\caption{ Illustrating the sample creation in a sub-region of the parameter space iterating from coarse to finer sampling. \mbox{(Un-)}filled circles indicate parameter configurations that lead to an \mbox{(un-)stable} steady state. \label{fig:stabcorner} }
\end{figure}

%\para{Enhancing detail views} \\
\para{Investigating pattern formation hypothesis} 
%\sbnote{
%Problem: there seems to be a relationship between potential to form patterns and the stability of the steady states of the system. However, the stability type <--> pattern formation potential relationship is still hypothetical and deserves further study and verification. \\
%Method: new detail views can be constructed on the fly. So, it only took us 10 minutes to glue the pattern display together with a display of a bifurcation diagram. }
To investigate the hypothesized relationship between pattern formation and linear stability analysis the customizable detail view feature proved helpful.
The main output shows a full spatio-temporal pattern as given in \FG{fig:gui}g.
%, but as mentioned in the previous paragraph their computation is optional.
%
Another view created for this application is a bifurcation diagram 
%\bluenote{would be nice to have brief explanation for that on ... it's not (yet) super common in Vis right? It's more a math thingy ... } \sbnote{see rest of the sentence.}
that shows how the multiplicity and stability type of all steady states (spatially homogenous solutions of the PDE system) are changing, as one parameter ($q_{al}$ in this case) is changing its value. This enables further study of possible relationships between steady states and pattern formation, as shown in the supplementary video material.
%For now, that is at \url{http://www.cs.sfu.ca/~sbergner/personal/proj/highd/paraglideswarms.html}.

%To allow for interactive exploration of these diagrams, their computation and display have been split into separate steps.

\para{Comparison of different model versions} 
%\sbnote{
%Problem: Comparison of different model versions (for constant and non-constant velocities) \\
%Method: rather than comparing the patterns one by one, the stability type feature extractor allows to visualize unstable regions (e.g. the white outside in Fig. 6) and a model comparison could be done by just showing how the regions change when switching model versions. \\
%Insight: instability tends to increase in the presence of non-constant velocities.}
%
The comparison of model versions using non-/constant velocities was enabled by creating two different feature variables computing the stability type using either of the two conditions. Switching the colour coding between these two variables allowed to visually compare the stability regions of the two model versions. 
This facilitated a main insight of the Master's thesis, showing that the instability of most steady states tends to increase in the presence of non-constant velocities.

Overall, the users in this case found \sname{Paraglide} to be: \textit{``a user-friendly tool that makes creating the sample points and comparing the computations much easier. This tool is capable of giving the user a better intuition about different solutions that correspond to various parameter configurations.''}

%% --------------------------------------------------------------------------
\subsection{Bio-medical imaging: Tuning Image Segmentation Parameters \label{sec:biomedres} }

%The segmentation algorithm assessment of \SC{sec:biomed} provides the setting for the following discussion of a workflow for multi-factor assessment for optimization of multiple objectives.

In the following, we evaluate the use of \sname{paraglide} in the context of \SC{sec:biomed}. 
During three recorded meetings of overall $6$ hours a workflow was developed, implemented, and the required interaction steps performed.
%Its description is interspersed with notes on how particular aspects of \sname{paraglide}'s design affected the work of the algorithm researcher.
The goal there was to find a robust setting for eight parameters of a segmentation algorithm that produces \defn{good} results for different data sets and noise levels, assessed by ten numeric objective measures.
The term {\em good} in the sense of this discussion, refers to all points on a plateau of the optimization landscape that have target values close to the global optimum. When chosen from an initial, explorative sample they are also referred to as candidates, or representatives of the good cluster.
Since {\em the} optimum is an ambiguous term in the context multi-objective optimization, our method proceeds by first grouping all points that are similar to each other. This leaves the task of finding out which cluster of points is a good one. The shape of the plateau of good points viewed in the space of input parameters informed the developer about which parameters to keep and which ones to drop. It also leads to a choice of configuration for the algorithm.
%A cluster of good points is identified from an initial sample of differently configured runs of the algorithm. Viewing the shape of this good cluster with respect to the input parameters allows to pick good configurations that are also robust in the other factors, as finally validated for different noise levels and patients.
%
To enable faster computation %while screening parameter settings, 
the volumetric patient data was reduced to a single slice that contains representatives from each class.

%\subsubsection{Optimizing parameter settings that are robust for different noise levels}
\para{Find good candidate points by visual inspection} 
%The initial sample is obtained by specifying a bounding region for all parameters and filling it with a sample of uniformly distributed points by using a Latin-hypercube based method.
%
For easier inspection, the full set of variables is first broken down into groups. 
A SPloM view of the input parameters verified that the sampling pattern indeed uniformly covers the 2D scatterplot projections. To focus on the problem, the user isolated the group of performance measures described in \SC{sec:biomed}.
Manually chosen configuration points improved one or two performance criteria, 
% (two of the ten objective measures)
and allowed to verify basic data sanity in a linked data table view.
A combined manual optimization of the $2$ performance measures for each of the $5$ classes, however, would require to pay attention to simultaneous changes in $5$ scatterplots. %, possibly even taking detail views into account.
The developer considered this a {\em very difficult to infeasible task that needed to be simplified}.

%% [1:16:40]
%Some energy terms amount to regularization, which improves generalization of the classifier and robustness to noise. 

%Indicators for good solutions are already constructed: 
%Dice coefficient gives distance from training data, also the approximation error of a kinetic model is available.

\para{Construct the good neighbourhood} 
%% Continuity of goodness
For most points in parameter space,
a continuous change in the input parameters leads to a  continuous change in the segmentation algorithm's behaviour and the derived performance measures.
This means that for each good point, it is worthwhile to explore the neighbourhood around it to find additional good and better settings~\cite{McIntosh:2007:single}.
With the distinction of input and output dimensions it is possible to construct and combine different notions of neighbourhood around a point.
The dialog of \FG{fig:compnode} is used to combine performance measures using weights that equalize the dynamic ranges. 
%The cosine similarity between the feature vectors is used to derive coordinates for a 
This feature vector space is then viewed using the spectral embeddings described in \SC{sec:embed}.
\begin{figure}[t]
\centering
\includegraphics[width=8cm]{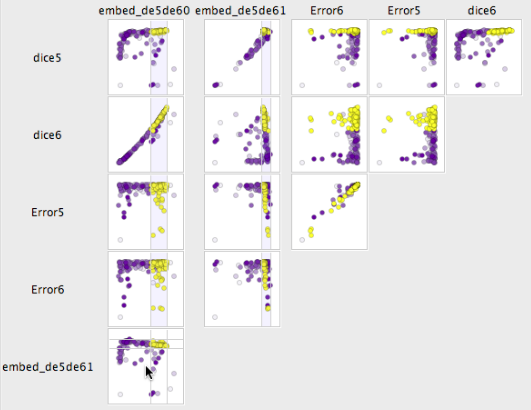}
\caption{Scatter plot matrix view that compares the point embedding (lower left) with the objective measures that went into computing its underlying similarity measure. The numbering of the responses corresponds to the class labels of \FG{fig:samedice}. 
%\ednote{TM: I don't think anyone understood your last sentence here. Can you be clearer?} 
\label{fig:EDembed}}
\end{figure}
\FG{fig:EDembed} shows the similarity embedding in the lower left view, where the good cluster is highlighted in yellow. %It was found by pointwise manual inspection and 
Judging from the strong diagonal distribution in two plots in the matrix, the horizontal embedding dimension is dominated by \cname{dice6} and the vertical one by \cname{dice5}. Since both should be maximized by good results, it is not surprising that manual inspection quickly identified the good cluster in the upper right of the embedding.
The user {\em found it convenient to make the cluster selection in the embedding}, which underlines the point of \SC{sec:beyondbox}. Apart from making interval selection easier, the embeddings also proved as an aid in a number of tasks:
  a) find good candidates, 
  b) group adjacent good points into cluster(s), and 
  c) check the embedding by inspecting it in a SPloM view together with the feature variables as in \FG{fig:EDembed}.

%This manual browsing and labelling of cluster representatives provides starting points for further refinement of the sample.

\para{Multi-factor assessment} 
To determine the relevance of each parameter for the overall performance of the segmentation algorithm, the developer viewed the distribution of the good cluster in input parameter space.
%As pointed out by Saltelli et al.~
%in \SC{sec:scatter},
This also gives a notion of sensitivity, 
%, where disconnected plateaus indicate a multi-modal target function and narrow peaks show optima that are sensitive to configuration changes.
where a large enough size of the good region indicates stability w.r.t. parameter changes~\cite[Sec. 1.2.3]{Saltelli:2008:gsa}. 
%One can pick the best point within it, or pick a number of good points spread out inside the region and choose their centre as best solution.

\begin{figure}[h]
\centering
\includegraphics[width=5cm]{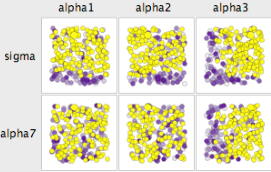}
\caption{Scatter plot matrix view of the good cluster (yellow) identified in \FG{fig:EDembed} viewed in the subspace of input parameters. 
While \cname{sigma} and \cname{alpha3} indicate clear thresholds beyond which the good configurations are found. 
%The other two appear as ``don't care'' parameters.
\label{fig:factorscreening}}
\end{figure}
When projecting the cluster onto each variable individually, its shape can be either spread out or localized in one or multiple density concentrations.
If the good points in the example of \FG{fig:factorscreening} are spread out along a dimension, the corresponding parameter is unusable for steering between good and bad performance, as in this case for ``don't care''  parameters \cname{alpha\{1,2,7\}}. 
Observations like that inform the developer of energy terms to drop and, hence, {\em directly influenced algorithm development}.
%Influential parameters are indicated, when the position of the good cluster is more localized or has a clear transition. 
Parameters showing more localized good points or a clear transition are kept as part of the segmentation model and are set to some robust value that is further from the boundary, inside the good region.

%For certain parameters the cluster might be multi-modal, exhibiting several disconnected density peaks. A possible step then is to refine the sampling further to get a clearer picture of the distribution around the peaks. The cluster component with the largest diameter indicates the most robust sub-region and is used to choose a setting for the corresponding parameter. \ednote{TM - this is directly contradicting your earlier statement that spreading out is bad! No it is good!?}

\para{Usage of the ROI representation}
%\sbnote{TODO: merge this with the list of section 4.3}
%In order to simplify the description of the good cluster during the assessment, the user constructed an enclosing region in input space, including the parameters shown in \FG{fig:factorscreening}. 
The region abstraction of \SC{sec:roi} found application simplifying several tasks:
  a) inspect its content by transferring it between projects considering different segmentation noise level or patient data, 
  %model version \ednote{TM: this seems to make no sense - you cannot use the same parameters in a different model, where they might have a totally different meaning or not even appear at all? } \sbnote{I'm using model version, not type, to refer to a differently configured energy function.}
  b) to adjust the region description under these different experimental conditions, 
  c) refine the sampling of configurations of the current model,
  d) communicate data requests via email.
The main steps of the user driven optimization perform (a) and (b) iteratively for runs with different noise levels. This results in a region description for good and robust parameter choices.
%With the indirect interface to the computation code, 
Refinement (c) was performed implicitly by applying (a) to a pre-computed denser data sample of $10,000$ points, which yielded $23$ good configurations with a segmentation quality similar to \FG{fig:samedice}d.
The best chosen configuration %we've got was #8374 in the 10000 file which results in 
of $\text{\cname{dice6}}=0.8282$ and $\text{\cname{error6}}=0.0621$,
was also verified to be visually convincing.

% Put (multiple) bounds on parameters that show local concentration.

%If multiple data sets are involved in the construction it is possible to combine their good regions, such that a non-empty intersection or best compromise gives robust high quality performance.

\para{Verify generalization} 
While the optimum has been made robust by constructing it over different experimental conditions, its performance has to generalize well beyond the condition used during adjustment. 
Hence, a final verification is run using data from $10$ previously unseen patients. 
Compared to the best configuration, the $10$ validation data sets showed very good Dice coefficient ($\mu=0.781540,\sigma=0.06$) and excellent kinetic modelling parameter error ($\mu=0.062, \sigma=0.0001$) throughout.
%Dice:                             mean 0.781540	std 0.06040     median 0.784000
%Kinetic modeling parameter error: mean 0.062203	std 0.000121	    median 0.062183
%
This indicates that the configuration overall delivers high shape accuracy as well as low kinetic error. 
Two of the $10$ data sets yield just above average, yet acceptable, Dice coefficients, which {\em inspired a separate investigation}. % that will consider configurations adapted to different patient populations. 
This shows that the interaction steps suggested by \sname{paraglide} can accelerate and benefit daily practice in state-of-the art research.
%The success of the workflow described here indicates that the abstractions implemented in \sname{paraglide} benefit daily practise in state of the art research.

%Since only one best configuration of the segmentation model is required, a distinction of different clusters of outcomes seems unnecessary.
%optimize a particular segmentation model, 
%However, distinct optimal solution plateaus may differ in size, which indicates their sensitivity to parameter changes. 
%Robustness of an optimum to slight perturbation of the input (i.e. reliable segmentation for varying anatomy of scanned patients) trumps a slight optimality improvement.
% Using this information in the inspection addresses requirement \R8.
%
%Being able to observe causes of different types of deviations from the ideal segmentation provides hints on how to improve the classifier in a qualitative way to modify the method, and helps to assess the effectiveness of new energy terms. 

\begin{comment}
%% comparison with T-W:
While there is a scatter plot based method recently devised by Torsney-Weir et al.~\cite{Torsney-Weir:2011:tuner}, we will focus on an alternative approach that makes use of 
a) embeddings to find good points more easily, because similar ones are close together. This allows to group them to create a good cluster, and
b) linked views to show values of the feature vectors and allow to look at distribution w.r.t. the parameters.
\end{comment}

%% --------------------------------------------------------------------------
\subsection{Fuel cell stack prototyping \label{sec:cstackres} }

The following case was introduced in \SC{sec:cstack} and concerns the simulation of a fuel cell stack. The model by Chang et al.~\cite{Chang:2007:cellstack} depends on about $100$ input parameters and produces $43$ different plots of various physical quantities characterizing the behavior of the cell stack.
The parameters are structured into semantic groups describing different parts of the assembly. Further, the developer of the code has provided short description texts for the variables and their physical units. These parameter groups and descriptions are passed on through the \cname{ComputeNode} interface of \SC{sec:compnode} and appear in \sname{paraglide} as prepared variable groups and tool tips.

%In the current approach 
A parameter region of interest is chosen by the user, giving value ranges for the selected dimensions. All other parameters are kept at constant default values. \sname{Paraglide} interfaces with the simulation code via a network connection, allowing multiple instances of the simulator to compute output for the generated sample configurations distributed over several computers. When all experiments are computed, one can choose an output plot of interest. The corresponding plots of all experiments are collected and compared using the correlation measure and layout method described at the end of \SC{sec:embed}. 
%A layout algorithm then arranges the samples on the screen, such that spatial proximity represents similarity in experimental outcome with respect to the chosen plot.
%
Since spatial proximity in these embeddings represents similarity in experimental outcome with respect to the chosen plot, {\em detail inspection and manual labelling of multi-dimensional clusters} using simple rectangle selection become feasible and improve confidence in the resulting decomposition.
%Again, experiments that exhibit distinct behaviour can be manually inspected and layout in the similarity embedding simplifies the clustering task to a degree that it is possible to manually label points and their similar neighbours using rectangle selection in the embedding.
%Further, the user may manually label experiments that exhibit distinct behavior. Currently, selection can be made by visual inspection and drawing rectangles on the screen. 

% -----------------------------------------------------------------------
% pre-include for proper page layout
%\clearpage
\begin{figure*}[tp]
\centering
\begin{tabular}{lcr}
  \subfigure[x-stack current, y-in temp.]{\includegraphics[width=5.5cm]{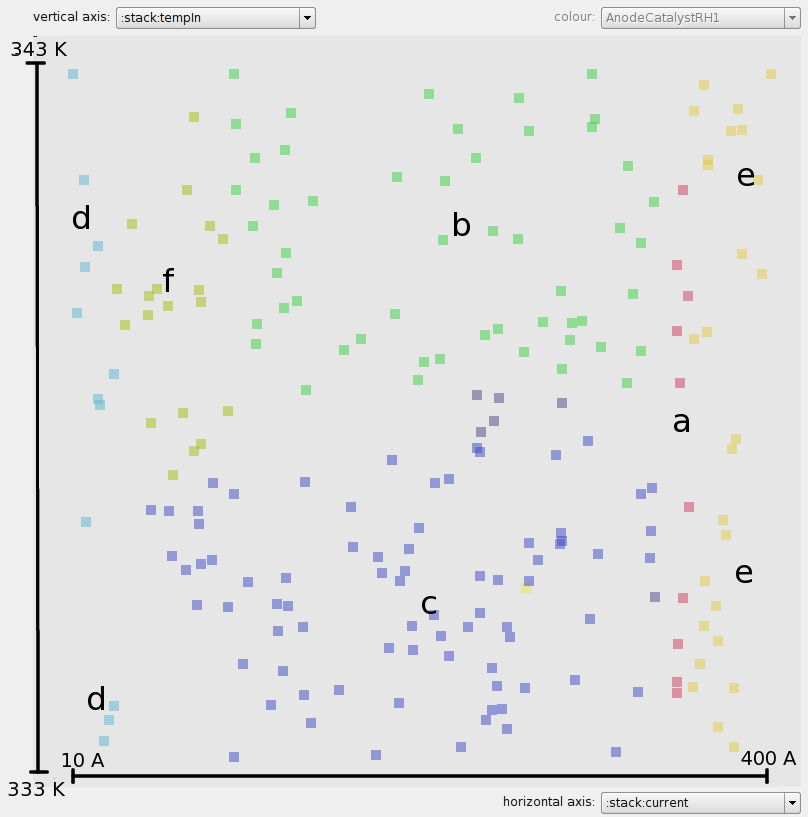}}
  & \subfigure[cell current density similarity]{\includegraphics[width=5.5cm]{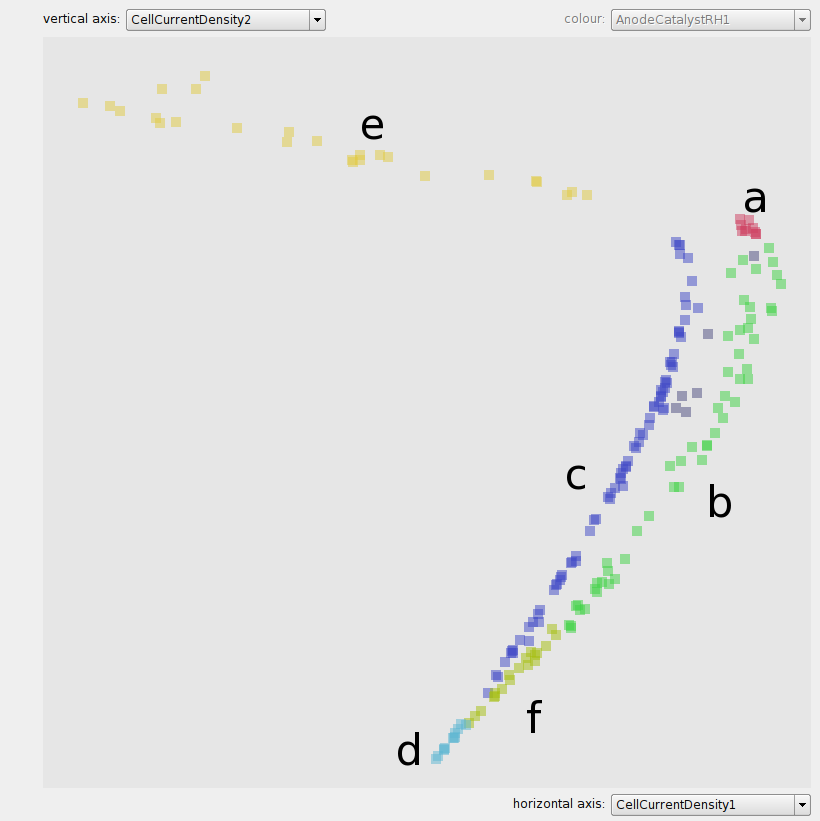}}
  & \subfigure[MEA water content similarity]{\includegraphics[width=5.5cm]{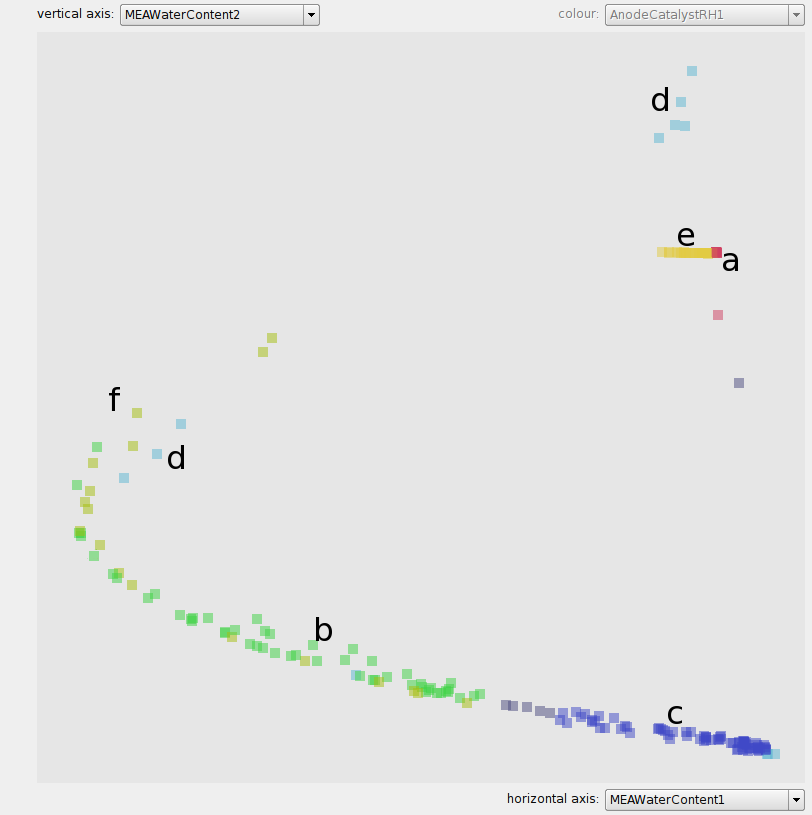}}
\end{tabular}
\caption{
  Two layouts for $204$ example experiments. a) input space
  showing variation in current and input temperature, b) embedding of
  the same samples where spatial proximity reflects plot
  similarity for cell current density, c) similarity embedding for membrane electrode assembly (MEA)
  water content using the same clusters as assigned in (b). Cluster representatives are
  shown in \FG{fig:plotsCCD} and \FG{fig:plotsH2O}.
  These screenshots are from the 2007 C++ version of \sname{paraglide}, and are also attainable in the currently discussed Java implementation. 
%\ednote{TM - you need to demonstrate the capabilities of the current tool.}
%\sbnote{These screenshots do reflect the capabilities of the current tool, because the algorithms and underlying ideas etc. used back then are exactly the same as the ones used now.}
  \label{fig:embed}}
\end{figure*}
\begin{figure}[p]
\centering
\begin{tabular}{lr}
  \subfigure[]{\includegraphics[width=4cm]{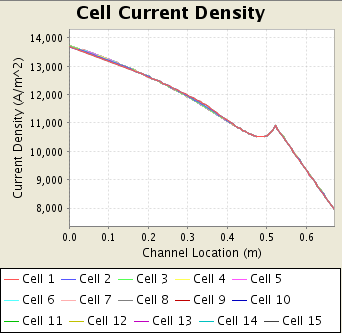}}
  & \subfigure[]{\includegraphics[width=4cm]{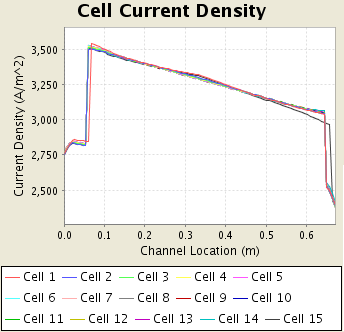}} \\
  \subfigure[]{\includegraphics[width=4cm]{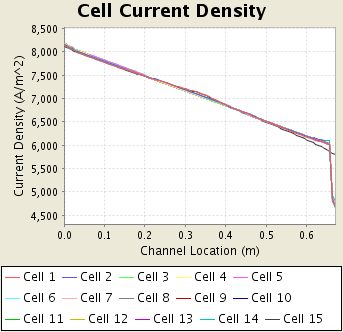}} 
  & \subfigure[]{\includegraphics[width=4cm]{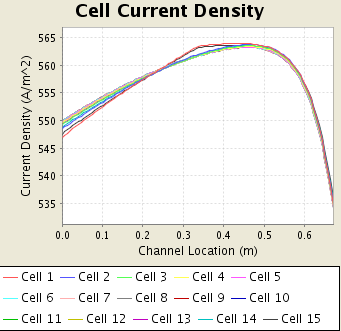}} \\
  \subfigure[]{\includegraphics[width=4cm]{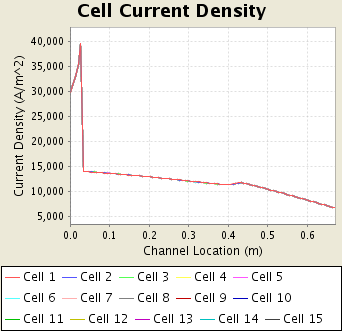}} 
  & \subfigure[]{\includegraphics[width=4cm]{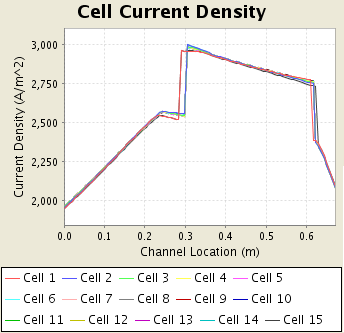}}
\end{tabular}
\caption{Cell current density plots for the clusters in \FG{fig:embed}b \label{fig:plotsCCD}}
\end{figure}
\begin{figure}[p]
\centering
\begin{tabular}{lr}
  \subfigure[]{\includegraphics[width=4cm]{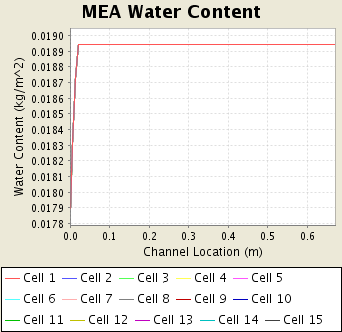}}
  & \subfigure[]{\includegraphics[width=4cm]{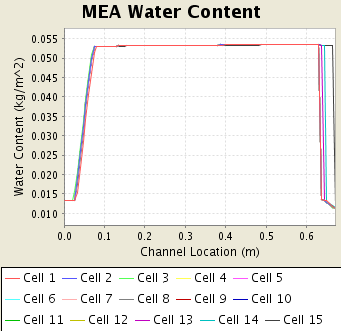}} \\
  \subfigure[]{\includegraphics[width=4cm]{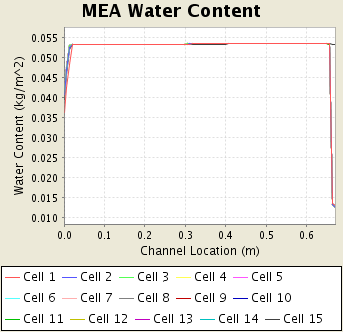}} 
  & \subfigure[]{\includegraphics[width=4cm]{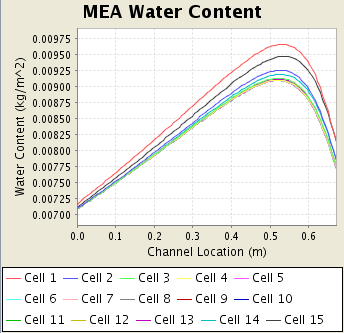}} \\
  \subfigure[]{\includegraphics[width=4cm]{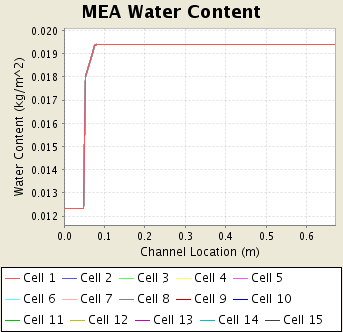}}
  & \subfigure[]{\includegraphics[width=4cm]{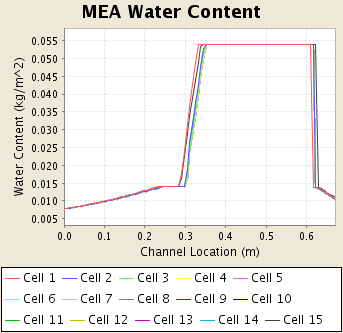}}
\end{tabular}
\caption{MEA water content plots for the clusters labelled in \FG{fig:embed}c \label{fig:plotsH2O}}
\end{figure}
%\clearpage

\para{Experiments with current and inflow temperature} 
To keep the initial experiment simple, we have chosen a region of
interest over two input parameters: stack current ($10A..400A$) and
stack inflow temperature ($333K..343K$). In this region $204$ samples
are created with a uniform random distribution shown in
\FG{fig:embed}a. The color coding is added at a later stage and has no
relevance for the initial step.

In \FG{fig:embed}b the sample configurations are arranged according to their 
similarity in cell current density. In this
embedding simulation outcomes can be inspected and configuration sample points can be manually
labeled using a screen rectangle selection. For comparison, another
similarity based embedding is shown in \FG{fig:embed}c for the water
content of the membrane-electrode assembly (MEA) using the same
cluster labels as \FG{fig:embed}b. When going back to the input space
\FG{fig:embed}a, the color coding reflects the parameter ranges of
distinct behavior.  Cluster representatives are shown in
\FG{fig:plotsCCD} and \FG{fig:plotsH2O}.

\begin{comment}
Value ranges identified when viewing the clusters from the above figures in input space:
%\begin{itemize}
 a) $T= 334..341 K$, $I= 347..354 A$
 b) $T= 338..343 K$, $I= 100..334 A$
 c) $T= 333..337 K$, $I= 53..333 A$
 d) $T= 333..343 K$, $I= 10..33 A$
 e) $T= 333..343 K$, $I= 356..400 A$
 f) $T= 337..342 K$, $I= 34..93 A$.
%\end{itemize}
\end{comment}

Our users found the parameter space partitioning of \FG{fig:embed} intriguing, giving them a new method to study their model. For instance, they pointed out that while cluster representative \FG{fig:plotsCCD}e may be physically unreasonable, 
% since extra adverse reactions should occur at this point, 
the (e) region in \FG{fig:embed}a can be interpreted as a ``bad'' region, where adverse reactions occur. 
\section{Conclusion and Future Work}

%\subsection{Lessons learned}

%A first version of \sname{paraglide} that was developed in 2006/2007 the focus was on 
%using embeddings for multi-dimensional filtering, and using continuous input/output relationship to improve scatterplot display.

The development of computer simulations needs careful setup of the involved parameters.
%To do this systematically, the effects have to be properly understood, and relevant criteria be formalized and combined into a single number that can be optimized analytically or numerically.
The discussed use cases indicated that the required understanding process can be time and resource intensive, and that systematic assistance is in order. Our validation of these investigations showed how the proposed decomposition of the continuous input parameter space benefits different questions. 
% of a computer simulation based on similarity in outcome. 
One finding suggests that even with potentially large numbers of variables, such a quantization can lead to a small number of regions.
%--- as for example the $6$ regions in the $100$ dimensional example of \FG{fig:embed}.
This can lead to a significant conceptual simplification and narrow down questions to particular sub-regions. 
Also, it provides starting points for local sensitivity analyses. % that investigate transitions between regions.

%The careful choice of abstractions and their flexible combination using scripting lead to a system that proved to be flexible and effective in a number of different application settings. For instance, the workflow described in \SC{sec:biomedres} was developed, implemented, and applied in the course of three meetings that together took less than 6 hours. 

%As for the overall implementation, the choice of \sname{prefuse} for rendering and control turned out to be a challenge. Producing histograms and scatter plots for many points in many linked views is expensive, as for instance selection update events have to be communicated through several cascaded tables. The ability to trigger data update propagation more selectively, would help to balance performance in larger-scale settings. Currently, this limits us to a maximum of about $1000$ points that can be displayed in subspaces of up to $10$ variables.
%Also, the inclusion of posted bugfixes would help  is behind.
%Very useful in our setting was the relational table interface with tuple expressions and an overall smooth integration with Java Swing.
%The region abstraction was implemented as a boolean combination of multiple RangeQueryBindings, which could be controlled by different GUI widgets.
%
%All plug-in code and the sampling, embedding, clustering modules are implemented in the scripting layer. This makes the multi-dimensional data visualization back-end replaceable and integration with other systems possible. 

\para{Future directions}
%There are numerous questions that deserve further study.
%
Feasible or interesting regions for large numbers of variables can be relatively small in volume when compared to their bounding box. This is an example setting, where domain specific hints can help to guide the choice of sample points.
For that, one could seek to obtain an implicit function representation of the region boundary, that could be used by the sampling module of \FG{fig:conview}.

%Multi-dimensional viewing and navigation techniques could be expanded to include continuous regions. In general, all techniques for viewing multi-dimensional (indicator-) functions fit in this context.
The region setup that has been identified as a requirement in several tasks, could consider navigation widgets for multi-dimensional parameter spaces. The aspects that have been investigated so far \cite{Bergner:2005:PSVR,U-SFraser-CMPT-TR:2011-4} lead to interesting combinations of user interaction and dimension reduction.

%What's available:
%- projection techniques:
%  parameterization (choice of suitable derived variables as alternative inputs or outputs),
%%  reconstruction of continuous density from a discrete sample
The current implementation facilitates different point grouping methods: 
manual labelling, application of a classifier function, 
and clustering via a user determined similarity measure. 
Since clustering is key to parameter space partitioning, further research on suitable interactive and (semi-)automatic techniques for this purpose would be useful.

The investigated decomposition method results in a set of continuous regions. Considering this, initial research 
%on this technique 
focussed on high-quality projection techniques for scatter plots of this kind of data. After adjusting research focus to sampling aspects and the use case evaluation presented here, suitable options to visually represent a region decomposition of a multi-dimensional continuous space still deserve further study.

\bibliographystyle{abbrv}
%%use following if all content of bibtex file should be shown
%\nocite{*}
\bibliography{paraglide,highd}

\begin{thebibliography}{10}

\bibitem{Berger:2011:mvpred}
W.~Berger, H.~Piringer, P.~Filzmoser, and E.~Gr{\"o}ller.
\newblock Uncertainty-aware exploration of continuous parameter spaces using
  multivariate prediction.
\newblock {\em Computer Graphics Forum}, 30(3):911--920, 2011.

\bibitem{U-SFraser-CMPT-TR:2011-4}
S.~Bergner, M.~Crider, A.~E. Kirkpatrick, and T.~M{\"o}ller.
\newblock Mixing board versus mouse interaction in value adjustment tasks.
\newblock Technical Report TR 2011-4, School of Computing Science, Simon Fraser
  University, Burnaby, BC, Canada, Sep 2011.

\bibitem{Bergner:2005:PSVR}
S.~Bergner, T.~M{\"o}ller, M.~Tory, and M.~S. Drew.
\newblock A practical approach to spectral volume rendering.
\newblock {\em IEEE Trans. on Vis. and Comp. Graphics}, 11(2):207--216,
  March/April 2005.

\bibitem{Bhatt:1995:paraparti}
V.~Bhatt and J.~Koechling.
\newblock Partitioning the parameter space according to different behaviors
  during three-dimensional impacts.
\newblock {\em Journal of applied mechanics}, 62:740, 1995.

\bibitem{Bishop:2006:PRML}
C.~M. Bishop.
\newblock {\em Pattern Recognition and Machine Learning}.
\newblock Springer, August 2006.

\bibitem{Brochu:2007:apl}
E.~Brochu, N.~D. Freitas, and A.~Ghosh.
\newblock Active preference learning with discrete choice data.
\newblock In J.~Platt, D.~Koller, Y.~Singer, and S.~Roweis, editors, {\em
  Advances in Neural Information Processing Systems 20}, pages 409--416. MIT
  Press, Cambridge, MA, 2008.

\bibitem{Buhl:2006:disorder}
J.~Buhl, D.~Sumpter, I.~Couzin, J.~Hale, E.~Despland, E.~Miller, and
  S.~Simpson.
\newblock From disorder to order in marching locusts.
\newblock {\em Science}, 312(5778):1402, 2006.

\bibitem{Chang:2007:cellstack}
P.~Chang, G.-S. Kim, K.~Promislow, and B.~Wetton.
\newblock Reduced dimensional computational models of polymer electrolyte
  membrane fuel cell stacks.
\newblock {\em J. Comput. Phys.}, 223(2):797--821, 2007.

\bibitem{Dice:1945:dist}
L.~R. Dice.
\newblock Measures of the amount of ecologic association between species.
\newblock {\em Ecology}, 26(3):297--Ð302, 1945.

\bibitem{Doleisch:2003:FDL}
H.~Doleisch, M.~Gasser, and H.~Hauser.
\newblock {Interactive feature specification for focus+ context visualization
  of complex simulation data}.
\newblock In {\em Proc. of the symposium on data visualisation 2003}, pages
  239--248. Eurographics Association, 2003.

\bibitem{DeOliveira:2003:expomine}
M.~Ferreira~de Oliveira and H.~Levkowitz.
\newblock {From visual data exploration to visual data mining: A survey}.
\newblock {\em IEEE Trans. on Visualization and Computer Graphics},
  9(3):378--394, 2003.

\bibitem{Fetecau:2009:biogroups}
R.~C. Fetecau and R.~Eftimie.
\newblock An investigation of a nonlocal hyperbolic model for self-organization
  of biological groups.
\newblock {\em J. Math. Biol.}, 61(4):545--579, 2010.

\bibitem{Francois:2007:highd}
D.~Francois.
\newblock {\em High-dimensional data analysis: optimal metrics and feature
  selection}.
\newblock PhD thesis, Universit{\'e} catholique de Louvain, 2007.

\bibitem{Grady:2006:RW}
L.~Grady.
\newblock Random walks for image segmentation.
\newblock {\em IEEE Trans. on Pattern Analysis and Mach. Intelligence},
  28(11):1768--1783, Nov. 2006.

\bibitem{Greenacre:1987:biplotca}
M.~Greenacre and T.~Hastie.
\newblock The geometric interpretation of correspondence analysis.
\newblock {\em J. of the Am. Stat. Assoc.}, 82(398):437--447, June 1987.

\bibitem{Heer:2006:DP}
J.~Heer and M.~Agrawala.
\newblock {Software design patterns for information visualization}.
\newblock {\em IEEE Trans. on Visualization and Computer Graphics},
  12(5):853--860, 2006.

\bibitem{Holbrey:2006:drvis}
R.~Holbrey.
\newblock Data reduction algorithms for data mining and visualization.
\newblock Technical report, University of Leeds/Edinburgh, 2006.

\bibitem{Ingram:2009:glimmer}
S.~Ingram, T.~Munzner, and M.~Olano.
\newblock {Glimmer: Multilevel MDS on the GPU}.
\newblock {\em IEEE Trans. on Visualization and Computer Graphics}, pages
  249--261, 2009.

\bibitem{Jaenicke:2008:brushing}
H.~J{\"a}nicke, M.~B{\"o}ttinger, and G.~Scheuermann.
\newblock {Brushing of Attribute Clouds for the Visualization of Multivariate
  Data}.
\newblock {\em IEEE Trans. on Visualization and Computer Graphics},
  14(6):1459--1466, 2008.

\bibitem{Jones:1998:ego}
D.~Jones, M.~Schonlau, and W.~Welch.
\newblock {Efficient global optimization of expensive black-box functions}.
\newblock {\em Journal of Global Optimization}, 13(4):455--492, 1998.

\bibitem{Kilian:2007:shapespace}
M.~Kilian, N.~J. Mitra, and H.~Pottmann.
\newblock Geometric modeling in shape space.
\newblock {\em ACM Transactions on Graphics}, 26(3):1--8, 2007.

\bibitem{Lemieux:2009:MCQMC}
C.~Lemieux.
\newblock {\em Monte Carlo and Quasi-Monte Carlo Sampling}.
\newblock Springer Verlag, 2009.

\bibitem{Lutscher:Stevens}
F.~Lutscher and A.~Stevens.
\newblock Emerging patterns in a hyperbolic model for locally interacting cell
  systems.
\newblock {\em J. Nonlinear Sci.}, 12:619--640, 2002.

\bibitem{Luxburg:2007:spectutor}
U.~Luxburg.
\newblock A tutorial on spectral clustering.
\newblock {\em Statistics and Computing}, 17(4):395--416, 2007.

\bibitem{Marks:1997:galleries}
J.~Marks, B.~Andalman, P.~A. Beardsley, W.~Freeman, S.~Gibson, J.~Hodgins,
  T.~Kang, B.~Mirtich, H.~Pfister, W.~Ruml, K.~Ryall, J.~Seims, and S.~Shieber.
\newblock Design galleries: {A} general approach to setting parameters for
  computer graphics and animation.
\newblock In {\em SIGGRAPH '97}, pages 389--400. ACM Press, 1997.

\bibitem{Martin:1995:hdbrush}
A.~Martin and M.~Ward.
\newblock High dimensional brushing for interactive exploration of multivariate
  data.
\newblock In {\em Proc. of 6th Conference on Visualization '95}, pages
  271--278. IEEE Computer Society, 1995.

\bibitem{McIntosh:2007:single}
C.~Mcintosh and G.~Hamarneh.
\newblock {Is a single energy functional sufficient? Adaptive energy
  functionals and automatic initialization}.
\newblock {\em Proc. MICCAI, Part II}, 4792:503--Ð510, 2007.

\bibitem{Mulder:1999:compsteer}
J.~Mulder, J.~van Wijk, and R.~van Liere.
\newblock A survey of computational steering environments.
\newblock {\em Future Generation Computer Systems}, 15(1):119 -- 129, 1999.

\bibitem{Parrish:ube}
J.~K. Parrish.
\newblock Using behavior and ecology to exploit schooling fishes.
\newblock {\em Environ. Biol. Fish.}, 55:157--181, 1999.

\bibitem{Pitt:2006:paraparti}
M.~Pitt, W.~Kim, D.~Navarro, and J.~Myung.
\newblock Global model analysis by parameter space partitioning.
\newblock {\em Psychological Review}, 113(1):57, 2006.

\bibitem{Pretorius:2011:paraspace}
A.~Pretorius, M.~Bray, A.~Carpenter, and R.~Ruddle.
\newblock Visualization of parameter space for image analysis.
\newblock {\em IEEE Trans. on Vis. and comp. Graph.}, page In Press., 2011.

\bibitem{Saad:2008:kinetic}
A.~Saad, G.~Hamarneh, T.~M{\"o}ller, and B.~Smith.
\newblock {Kinetic modeling based probabilistic segmentation for molecular
  images}.
\newblock {\em Medical Image Computing and Computer-Assisted
  Intervention--MICCAI 2008}, pages 244--252, 2008.

\bibitem{Saltelli:2008:gsa}
A.~Saltelli, M.~Ratto, T.~Andres, F.~Campolongo, J.~Cariboni, D.~Gatelli,
  M.~Saisana, and S.~Tarantola.
\newblock {\em {Global Sensitivity Analysis: The Primer}}.
\newblock Wiley-Interscience, 2008.

\bibitem{Santner:2003:DACE}
T.~Santner, B.~Williams, and W.~Notz.
\newblock {\em The Design and Analysis of Computer Experiments}.
\newblock Springer, 2003.

\bibitem{Smith:2007:navishape}
R.~Smith, R.~Pawlicki, I.~K{\'o}kai, J.~Finger, and T.~Vetter.
\newblock Navigating in a shape space of registered models.
\newblock {\em IEEE Trans. on Vis. and Comp. Graphics}, pages 1552--1559, 2007.

\bibitem{Torsney-Weir:2011:tuner}
T.~Torsney-Weir, A.~Saad, T.~M{\"o}ller, B.~Weber, H.-C. Hege, J.-M. Verbavatz,
  and S.~Bergner.
\newblock {Tuner: Principled Parameter Finding for Image Segmentation
  Algorithms Using Visual Response Surface Exploration}.
\newblock {\em IEEE Trans. on Vis. and Comp. Graphics}, pages ???--???, 2011.

\bibitem{Trefethen:1997:NLA}
L.~Trefethen and D.~B. III.
\newblock {\em Numerical Linear Algebra}.
\newblock SIAM, 1997.

\bibitem{Tweedie:1998:prosection}
L.~Tweedie and R.~Spence.
\newblock The prosection matrix: A tool to support the interactive exploration
  of statistical models and data.
\newblock {\em Computational Statistics}, 13(1):65--76, 1998.

\bibitem{van:1993:hyperslice}
J.~van Wijk and R.~van Liere.
\newblock {HyperSlice: Visualization of scalar functions of many variables}.
\newblock In {\em Proc. of 4th Conf. on Visualization'93}, pages 119--125. IEEE
  Computer Society, 1993.

\bibitem{Wijk:1997:CSE}
J.~van Wijk, R.~Van~Liere, and J.~Mulder.
\newblock {Bringing computational steering to the user}.
\newblock In {\em Scientific Visualization Conference, 1997}, pages 304--304.
  IEEE, 1997.

\bibitem{Wong:1994:mdmv}
P.~Wong and R.~Bergeron.
\newblock {Years of Multidimensional Multivariate Visualization}.
\newblock {\em Scientific Visualization, Overviews, Methodologies, and
  Techniques}, pages 3--33, 1994.

\end{thebibliography}

\end{document}